\documentclass[aps,prd,twocolumn,amsmath,amssymb,showpacs,floatfix,superscriptaddress,nofootinbib]{revtex4-1}

\usepackage{natbib}
\usepackage{amsmath}
\usepackage{amssymb}
\usepackage{latexsym}
\usepackage{bm}   
\usepackage{graphicx,epstopdf}
\usepackage{float}
\usepackage{epstopdf}
\DeclareGraphicsExtensions{.ps}
\usepackage{dcolumn} 
\usepackage{multirow}
\usepackage{appendix}
\usepackage{footnote}
\usepackage{tabularx,ragged2e,booktabs}
\usepackage{xcolor}
\usepackage{pdfcomment}
\usepackage{simplewick}

\newcolumntype{C}[1]{>{\Centering}m{#1}}

\usepackage{color}

\DeclareMathAlphabet\mathbfcal{OMS}{cmsy}{b}{n}
\definecolor{darkgreen}{cmyk}{0.85,0.2,1.00,0.35} 
\definecolor{purple}{cmyk}{0.5,1.0,0,0} 
\definecolor{darkblue}{cmyk}{1.0,1.0,0,0}
\usepackage{hyperref}

\hypersetup{
   colorlinks = true,
   allcolors = darkblue}

\def\beq{\begin{equation}}
\def\eeq{\end{equation}}
\def\bea{\begin{eqnarray}}
\def\eea{\end{eqnarray}}
\def\lsim{\mathrel{\raise.3ex\hbox{$<$\kern-.75em\lower1ex\hbox{$\sim$}}}}
\def\gsim{\mathrel{\raise.3ex\hbox{$>$\kern-.75em\lower1ex\hbox{$\sim$}}}}
\def\wigner#1#2#3#4#5#6{ \left( \begin{array}{ccc} #1 & #3 & #5
\\ #2 & #4 & #6 \\ \end{array} \right)}

\newcommand{\TT}{_L^{TT}}
\newcommand{\TE}{_L^{TE}}
\newcommand{\EE}{_L^{EE}}

\def\hn{\hat{\mathbf{n}}}

\def\hD{\hat{\Delta}}
\def\LM{_{LM}}
\def\lm{_{lm}}

\def\TT{^{TT}}
\def\TE{^{TE}}
\def\EE{^{EE}}

\def\TB{^{TB}}
\def\EB{^{EB}}
\def\cc{^{*}}

\newcommand{\bn}{\hat{\bm n}}

\begin{document}
	
\title{Lensing Bias to CMB Polarization Measurements of Compensated Isocurvature Perturbations}

\author{Chen Heinrich}\email{chenhe@uchicago.edu}
\affiliation{Department of Physics, University of Chicago, Chicago, IL 60637}
\affiliation{Kavli Institute for Cosmological Physics, Enrico Fermi Institute, Chicago, IL 60637}

\begin{abstract}

Compensated isocurvature perturbations are opposite spatial fluctuations in the baryon and dark matter (DM) densities. They arise in the curvaton model and some models of baryogenesis. While the gravitational effects of baryon fluctuations are compensated by those of DM, leaving no observable impacts on the cosmic microwave background (CMB) at first order, they modulate the sound horizon at recombination, thereby correlating CMB anisotropies at different multipoles. As a result, CIPs can be reconstructed using quadratic estimators similarly to CMB detection of gravitational lensing. Because of these similarities, however, the CIP estimators are biased with lensing contributions that must be subtracted. These lensing contributions for CMB polarization measurement of CIPs are found to roughly triple the noise power of the total CIP estimator on large scales. In addition, the cross power with temperature and $E$-mode polarization are contaminated by lensing-ISW (integrated Sachs-Wolfe) correlations and reionization-lensing correlations respectively. For a cosmic-variance-limited (CVL) temperature and polarization experiment measuring out to multipoles $l_{\max} = 2500$, the lensing noise raises the detection threshold by a factor of 1.5, leaving a $2.7\sigma$ detection possible for the maximal CIP signal in the curvaton model.

\end{abstract}
\maketitle

\section{Introduction}

Measurements of the cosmic microwave background (CMB) have shown that the primordial perturbations in the Universe are mainly adiabatic~\cite{Ade:2013uln,Ade:2015lrj,Ade:2015ava}. These adiabatic perturbations are representative of single-field inflation, which gives all particle species the same fractional spatial fluctuations in their number density. On the other hand, the isocurvature perturbations arise as the difference between the fractional perturbations of two species, indicative of additional fields during inflation~\cite{Bond:1984fp,Kodama:1986fg,Kodama:1986ud,Hu:1994jd,Moodley:2004nz,Bean:2006qz}. In particular, the effective matter to radiation isocurvature mode has been highly constrained by the Planck mission to be less than a few percent of the adiabatic mode~\cite{Ade:2015lrj}. Here the effective matter refers to the combined effect of cold dark matter (CDM) and baryon fluctuations weighted according to their energy density. 

There is, however, one special class of perturbations that escapes the effective matter constraint, the compensated isocurvature perturbations (CIPs). In the CIP mode, the CDM and baryon density fluctuations are opposite of each other, giving no net gravitational effects, and hence no effective matter or radiation perturbations~\cite{Lewis:2002nc,cambnotes,Holder:2009gd,Gordon:2009wx,Ade:2015ava}. CIPs are therefore orthogonal to the effective matter isocurvature, and evade CMB constraints on these modes.

CIPs naturally arise in the curvaton model, as well as some models of baryogenesis~{\cite{DeSimone:2016ofp}. In the curvaton model, an additional scalar field during inflation -- the curvaton -- generates most of the adiabatic perturbations in lieu of the inflaton~\cite{Lyth:2002my,Gupta:2003jc,Gordon:2002gv}. Depending on different scenarios, i.e.,  the epochs when CDM and baryon number are created relative to the curvaton decay, there would be different amounts of CIPs produced. Those CIPs from the curvaton model would always be correlated with the adiabatic perturbations~\cite{Lyth:2001nq,Lyth:2002my}, and full correlation happens if the curvaton contribution to adiabatic perturbations is dominant over the inflaton. In the fully correlated case, the largest CIP has amplitude relative to the adiabatic perturbations $A \approx 16.5$, within the reach of the next generation of nearly cosmic-variance-limited (CVL) CMB experiments~\cite{He:2015msa}.

CMB observations are a particularly clean probe for CIPs as they are not dependent on particular assumptions such as galaxy physics. In particular, quadratic estimators provide the optimal signal-to-noise for measuring CIPs with nearly CVL polarization experiments~\cite{Smith:2017ndr}. Even though CIPs leave no imprint in the CMB power spectra at first order (with compensating gravitational effects from CDM and baryons)~\cite{Lewis:2002nc,Gordon:2002gv}, the baryon density fluctuations still cause a modulation in the damping scale and sound speed of the baryon-photon fluid. As a result, the sound horizon at recombination varies spatially, breaking the statistical isotropy of the CMB. This variation correlates temperature and polarization anisotropies of different multipole moments, providing a way for us to reconstruct the CIPs using quadratic estimators~\cite{Grin:2011nk,Grin:2011tf}. 

Using the quadratic estimator technique, the authors in Ref.~\cite{He:2015msa} forecasted that a Stage-4 CMB experiment would be able to detect the maximal CIP scenario of the curvaton model at $3\sigma$. This sensitivity relies crucially on the use of nearly CVL polarization measurements at two steps: 1) in forming the total CIP estimator, by adding four more $E$ and $B$-mode based estimators to the $TT$ estimator, thereby reducing the estimator noise significantly~\cite{Grin:2011nk} and 2) in cross-correlating the reconstructed CIP map with $T$ and $E$-mode polarization, a crucial step that improves the sensitivity to correlated CIPs by a factor of 2 to 3~\cite{He:2015msa}.

The above forecast, however, does not include the effect of gravitational lensing which would also induce correlations between the different CMB multipoles~\cite{Okamoto:2003zw}. As the CIP estimators are designed to be unbiased for Gaussian CMB fields, the non-Gaussian CMB in the presence of lensing introduces a bias to the CIP estimators that must be removed with its error budget properly taken into account. In fact, the lensing bias properties have been simulated and studied for CIP measurements using CMB temperature alone, and was shown to degrade CIP detectability by a factor of 1.3~\cite{Heinrich:2016gqe}. A study of the lensing bias to CMB \emph{polarization} measurements of CIPs, however, has yet to be performed.

In this paper, we simulate the lensing bias to the total CIP estimator, composed of five single estimators  -- $TT, TE, EE, TB$ and $EB$ -- and evaluate its impact on the detectability of fully correlated CIPs for a CVL experiment. We find that the $B$-estimators $TB$ and $EB$ play a crucial role in reducing the lensing bias in the total estimator. They are the least contaminated and help reduce the bias on scales where they dominate the total estimator. To further exploit this fact, new optimal weights are derived directly from simulations, reducing the total estimator noise on scales $L\gtrsim40$.  

Despite the reduced bias on smaller scales, the noise power of the total estimator on large scales is still a factor of three higher than without lensing contamination. In the cross-spectrum with CMB $E$-mode polarization, we find a contamination coming from the large-scale correlation of reionization and lensing potential through the $TT, TE, EE$ estimators. In contrast, the $B$-estimators $TB$ and $EB$ do not reconstruct a strong lensing signal as their CIP signal dominates over CMB multipole pairs where the lensing signal is suppressed. Finally a similar contribution from lensing to integrated Sachs-Wolfe (ISW) correlation contaminates the total CIP-temperature cross spectrum, as was found for the $TT$ estimator in Ref.~\cite{Heinrich:2016gqe}. As a result of the lensing bias in all the CIP auto and cross spectra, the CVL detectability of correlated CIPs is reduced by factor of 1.5.

More specifically, we simulate 4000 realizations of lensed CMB temperature and polarization maps and compute the CIP reconstruction in position space, using efficient estimator forms given in the Appendix~\ref{sec:efficient_est}. We include no CIP signal in the maps so as to study the noise properties of the estimator. In order to isolate the non-Gaussian contributions of lensing, we also perform the same reconstruction on 4000 realizations of Gaussian CMB maps. We find that both with or without lensing, the noise in the total estimator can be treated to good approximation as Gaussian distributed, obeying a $\chi^2$ and Wishart distribution respectively in its auto power and cross power with other CMB fields. We also find no evidence for correlations between the noise power at different multipoles. The above properties guide the construction of the Fisher matrix used to forecast the final CIP detectability. 

This paper is divided as follows. We begin by reviewing the physics of CIPs
and the relevant curvaton scenarios in section~\ref{sec:background}.
In section~\ref{sec:simulations} we describe the simulations and the reconstruction
pipeline, and study the lensing contributions to the single and
total CIP estimator noise power spectra. In section~\ref{sec:forecasts}, we use Fisher matrix
technique to predict, for a CVL experiment, the degradation of CIP detectability when lensing bias is included.

\section{Background}\label{sec:background}

In this section we briefly review the physics of compensated isocurvature perturbations - their observable impacts on the CMB and how they originate from the curvaton model. We refer the reader to Refs.~\cite{Grin:2011tf,He:2015msa} for more details. \\

\subsection{Compensated isocurvature perturbations}\label{sec:cip}

Isocurvature perturbations are the differences between the fractional number density perturbations of different species. With respect to the photon perturbations, the isocurvature mode of a species $i \in \left\{b,c,\nu,\gamma\right\}$ is defined as
\begin{equation}
		S_{i\gamma}=\frac{\delta n_{i}}{n_{i}}-\frac{\delta n_{\gamma}}{n_{\gamma}}, 
\end{equation} 
where $b$ stands for baryons, $c$ for cold dark matter, $\nu$ for neutrinos, and $\gamma$ for photons.

Compensated isocurvature perturbations are a special type of isocurvature mode in which the baryon and dark matter density fluctuations cancel
\begin{eqnarray}
	S_{b\gamma} = \Delta ,\quad	S_{c\gamma} = -\frac{\rho_b}{\rho_c} \Delta, \quad
	S_{\nu\gamma} = 0.
\end{eqnarray}
As a result, the CIP mode does not contribute to the effective matter isocurvature $S_{m\gamma} \equiv S_{b\gamma} + (\rho_c/\rho_b) S_{c\gamma}$. 

Because the gravitational effects of baryons and CDM are compensated, the CIPs have no observable impacts on the CMB power spectrum at first order. However, its baryon perturbations lead to spatial fluctuations of the sound speed, affecting CMB acoustic modes. Only CIPs on scales larger than the sound horizon at recombination leave a significant imprint, otherwise the spatially modulating speed would average out over one or more wavelengths as the sound waves travel until recombination. For CIPs larger than the sound horizon, the effects on the CMB modes are modelled with a separate-universe (SU) approach as perturbations in the background densities 
	\begin{align} 
	\delta \Omega_b = \Omega_b \Delta, \quad \delta \Omega_c = - \Omega_b\Delta.
	\label{eqn:separate}
	\end{align}

At first order in CIP, there is no observable impact on the CMB anisotropy angular power spectra. These are calculated given the primordial curvature power spectrum $P_{\zeta\zeta}$ as
	\begin{align}
			C_l^{XY} = \frac{2}{\pi}\int k^{2}dk T_{l}^{X}(k) T_l^{Y}(k)P_{\zeta\zeta}(k), 
	\label{eqn:ClXY}
	\end{align}	
where $X, Y \in \{\tilde{T}, \tilde{E}\}$ are the unlensed CMB temperature and polarization fields, and $C_l^{\tilde{B}\tilde{B}} = 0$ as we assume no primordial tensor perturbations. 
	
We can Taylor expand to first order (as appropriate for small CIPs) the transfer functions that encode the dependence on background densities, and obtain the derivative power spectra as
	 \begin{align}
			C_l^{X,dY} = \frac{2}{\pi}\int k^{2}dk T_{l}^{\tilde{X}}(k) \frac{d T_l^{\tilde{Y}}}{d\Delta}(k)P_{\zeta\zeta}(k).
			\label{eqn:Clderiv}
	\end{align}
where $X, Y \in \{T, E\}$. In the absence of tensors the $B$-mode derivative power spectra start only at second order. In this calculation, we expand upon the unlensed rather than the lensed CMB because we are modelling the CIP effects at the surface of last scattering, where gravitational lensing by large scale structure have not yet occurred.

As a three-dimensional field however, CIPs also affect the process of reionization at a later redshift. If we ignored reionization effects in the transfer functions used to obtain the derivative power spectra, we would be conflating, during the CIP reconstruction, different $k$-modes contributions from the epochs of reionization and recombination to the same reconstructed multipole $L$. To avoid this problem, we roughly model the reionization signal by fixing the optical depth $\tau$, and allowing the baryon density to modulate the redshift of reionization. In reality, the spatial modulations of baryon and DM densities would also impact the details of nonlinear structure formation leading to reionization. However, a complete modelling of such a reionization signal from a three-dimensional CIP field is beyond the scope of this paper, so we simply focus on the approximate effect in the redshift of reionization.

Finally we decompose the CIPs at the surface of last scattering as
\beq
\Delta(\bn) = \sum_{LM} \Delta_{LM} Y_{LM},
\eeq
with $L\lesssim 100$ being the valid range of the SU approximation. We use quadratic reconstruction to recover each $\Delta_{LM}$ mode, similarly to CMB measurements of gravitational lensing. Because of these very similarities, the CIP measurements will be contaminated by the lensing signal, which we will study in detail throughout this paper.

\subsection{Curvaton} \label{sec:curvaton}

One possible physical origin of CIPs is the curvaton model. In this model, the curvaton -- a spectator scalar field during inflation -- is responsible for seeding most of the adiabatic perturbations in the Universe. It later decays and seeds isocurvature perturbations correlated with the adiabatic perturbations~\cite{Mollerach:1989hu,Linde:1996gt,Lyth:2001nq,Lyth:2003ip,Lemoine:2006sc}. 
In the different decay scenarios, baryon number and DM can be generated either as a product of the curvaton decay, non-thermally before the decay, or out of the thermal plasma after the decay. Depending on the scenario, the fractional perturbations in the species will be different, leading to correlated isocurvature perturbations, and in particular, to correlated CIPs. 

If all of the adiabatic perturbations come from curvaton contributions, the resulting CIPs will be fully correlated. We use $A$ to denote the relative amplitude to adiabatic perturbations for the fully correlated CIPs
    \begin{equation}
   \Delta = A \zeta.
    \end{equation}
Two scenarios have large enough CIPs measurable with upcoming CMB polarization experiments: $A \approx 3\Omega_c/\Omega_b \approx 16.5$ (baryon produced by curvaton decay and CDM before decay) and $A = -3$ (CDM by decay, and baryons before)~\cite{He:2015msa}. 

For these fully correlated CIPs, we can exploit the additional signal available in the cross-correlations with the CMB anisotropies which are themselves evolved out of the adiabatic perturbations according to Eq.~\ref{eqn:ClXY}. We calculate the CIP power spectra $C_l^{X\Delta}$ with $X \in T, E, \Delta$ using Eq.~\ref{eqn:ClXY} with $T_l^{Y}$ replaced by the CIP transfer function 
\begin{align}
T_l^\Delta(k) = A j_l(kD_*), 
\end{align}
which is basically a projection onto a spherical shell at the distance to recombination $D_*$ using Bessel functions $j_l$. 

For the signal calculation, we now use the lensed CMB fields $T$ and $E$ with reionization contributions included as would be the case for real CMB observations. In the relevant separate-universe limit on scales of $l\lesssim200$, the lensed and unlensed CMB differ negligibly. 
On the other hand, reionization effects dominate the $E$-mode signal for $l\lesssim20$. Since the CIP transfer function is only a projection at recombination and does not model reionization signals of CIPs, the $C_L^{E\Delta}$ calculation here is free of unwanted correlation from a reionization signal. 

The large-scale reionization signal does correlate, however, between the CIP reconstruction (see section.~\ref{sec:cip}) and the observed $E$-mode. In particular, through enhanced responses in $C_{l'}^{T,dE}$ and $C_{l'}^{E,dE}$, it lowers the $TB$ and $EB$ estimator noise compared to the expected scale-invariant spectrum at low-$L$. Since the unwanted correlation from large-angle reionization signal does not reflect the true correlation between CIPs and the adiabatic perturbations and would artificially enhance the detectability of correlated CIPs, we set $C_{l'}^{T,dE}$ and $C_{l'}^{E,dE}$ to zero for $l'\leq20$ in the CIP reconstruction of Sec.~\ref{sec:simulations}.

\section{Simulations}\label{sec:simulations}

In this section, we simulate CIP reconstruction from CMB temperature and polarization maps, and characterize the reconstruction noise properties with and without non-Gaussian contributions from CMB lensing. We work with a flat $\Lambda$CDM cosmology consistent with the Planck 2015 results \cite{Ade:2015xua} with baryon density $\Omega_b h^2$ = 0.02225, cold dark matter density $\Omega_c h^2$ = 0.1198, Hubble constant $h = 0.6727$, scalar amplitude $A_s = 2.207\times10^{-9}$, spectral index $n_s = 0.9645$,  reionization optical depth $\tau = 0.079$, one massive neutrino with $m_\nu \approx 0.06$eV, CMB temperature $T_{\rm cmb}$ = 2.726K and no primordial tensor perturbations. The lensing simulations are performed using CAMB\footnote{CAMB: \url{http://camb.info}}\cite{Lewis:1999bs},
LensPix\footnote{LensPix: \url{http://cosmologist.info/lenspix/}} \cite{Lewis:2005tp, Hamimeche:2008ai}, and HEALPix\footnote{HEALPix: \url{http://healpix.sourceforge.net}} \cite{Gorski:2004by} and 
a modified version of LensPix for the CIP reconstruction that we now describe. 

\subsection{CIP reconstruction} \label{sec:gaussian}

To test our reconstruction pipeline, we start with the case of Gaussian CMB fields, for which we can analytically predict the expected noise properties. Since we are only interested in the reconstruction noise, we take the amplitude of the CIP signal to be zero in all of our simulations. 

Using Lenspix, we draw independent unit Gaussian variates that linearly combine to form CMB multipoles $\hat{T}_{lm}, \hat{E}_{lm}, \hat{B}_{lm}$ for $n_{\rm sim} = 4000$ realizations. We use the Cholesky decomposition of the covariance matrix~\cite{Lewis:2011fk} so the correlations are consistent with the lensed power spectra $C_l^{TT}$, $C_l^{EE}$, $C_l^{BB}$ and $C_l^{TE}$ (by parity $C_l^{TB}$ = $C_l^{EB}$ = 0), and call these Gaussian CMB maps for short. Note that in the absence of tensor perturbations $C_l^{BB}$ arises purely from the gravitational lensing of $E$-modes. Furthermore, these maps do not contain any non-Gaussian correlations that a proper lensing procedure of pixel-remapping would produce. For these simulations, we have chosen $N_{\rm side} = 2048$ and $l_{\rm max} = 3900$, and verified that these settings are sufficient for accurately evaluating estimators with modes to $l_{\rm est, max} = 2500$.

Next, we compute single CIP estimators using quadratic pairs $XZ$ of the CMB temperature and polarization fields. The harmonic-space form of the minimum-variance estimators is~\cite{He:2015msa,Grin:2011tf}
\beq
\hat{\Delta}_{LM}^{XZ}= N_{L}^{XZ} \sum_{lm l'm'}X_{l'm'}^{*}Z_{lm}g_{L l l'}^{XZ, \rm mv}\xi^{LM}_{lm l'm'},\label{eq:single_est}\\
\eeq
where $XZ \in \{TT, TE, EE, TB, EB\}$, 

\beq
\left[N_{L}^{XZ}\right]^{-1}=\sum_{l l'}G_{l l'}S_{l l'}^{L,XZ}g_{L l l'}^{ XZ, \rm mv}\label{eq:single_norm}
\eeq
is the normalization required for an unbiased estimator in the absence of lensing, 
\begin{align}		
    			 \xi^{LM}_{l m l' m'} 		
			= & \,	(-1)^{m}		\sqrt{\frac{(2L+1)(2l+1)(2l'+1)}{4\pi}}	
\notag \\			
			& \times 	     \wigner{l}{-m}{L}{M}{l'}{m'},							
			\end{align}		
\beq
G_{l l'} = \frac{(2l+1)(2l'+1)}{4\pi},
\eeq
$S_{l l'}^{L,XZ}$ are response functions given by Table~\ref{tab:response}\footnote{We note a sign flip in $S_{l l'}^{L,EB}$ in front of the $C_l^{B,dB}$ term in Table II of Ref.~\cite{He:2015msa}. This term does not enter our calculations here as we do not consider tensors. 
},where
\begin{align}
    			_{s}H^{L}_{ll'}	\equiv &		\wigner{l}{s}{L}{0}{l'}{-s} \nonumber\\
\end{align}
are Wigner 3$j$ coefficients.

\begin{table}
\caption[CIP response functions]{The response function $S^{L, XZ}_{l l'}$ of the various two-point observables
 in Eq.~(\ref{eq:single_noise}).	} 
\begin{ruledtabular}
\centering
\label{tab:response} 
\begin{tabular}{c c c} 
$XZ$		& $S^{L, XZ}_{l l'}$ & $l + l' + L$\\	\noalign{\smallskip}
\hline \noalign{\smallskip}
$TT$			& $(C^{T,dT}_{l'} + C^{T,dT}_{l})\, _{0}H^{L}_{ll'}$& even\\ 	\noalign{\smallskip}
$TE$ 		& $C^{T,dE}_{l'} \,_{2}H^{L}_{ll'} + C^{E,dT}_{l} \,_{0}H^{L}_{ll'}$ & even \\ \noalign{\smallskip}
$EE$ 		& $(C^{E,dE}_{l'} + C^{E,dE}_{l}) \, _{2}H^{L}_{ll'}$& even \\ 	\noalign{\smallskip}
$TB$		& $-i C^{T,dE}_{l'} \,_{2}H^{L}_{ll'}$ & odd	 \\ \noalign{\smallskip}
$EB$		& $-i C^{E,dE}_{l'} \,_{2}H^{L}_{ll'}$ & odd \\	\noalign{\smallskip}
\end{tabular} 
\end{ruledtabular}
\end{table}

The weight functions that minimize the single estimator variance are given by
\beq
g_{L l l'}^{XZ, \rm mv}=\frac{S_{l l'}^{L,XZ*}C_{l}^{XX}C_{l'}^{ZZ}-\left(-1\right)^{l+l'+L}S_{l' l}^{L,XZ*}C_{l}^{XZ}C_{l'}^{XZ}}{C_{l'}^{XX}C_{l}^{ZZ}C_{l}^{XX}C_{l'}^{ZZ}-\left(C_{l}^{XZ}C_{l'}^{XZ}\right)^{2}}.
\label{eq:single_noise}
\eeq
where we use the lensed CMB power spectra.

In practice, we compute efficiently the single estimators as a product of two maps using the
position space expressions given in Appendix~\ref{sec:efficient_est}.
They are equivalent to the harmonic space forms above for all except the $TE$ estimator.
In the $TE$ case, the position space form can only be achieved if we dropped the second term in the denominator of the minimum-variance weight function $g_{L l l'}^{TE, \rm mv}$, so that for $TE$ only, we have instead
\beq
\bar{g}_{L l l'}^{XZ} = \frac{S_{l l'}^{L,XZ*}C_{l}^{XX}C_{l'}^{ZZ}
-\left(-1\right)^{l+l'+L}S_{l' l}^{L,XZ*}C_{l}^{XZ}C_{l'}^{XZ}}
{C_{l'}^{XX}C_{l}^{ZZ}C_{l}^{XX}C_{l'}^{ZZ}}
\eeq
in the sum as well as in the normalization for unbiasedness. As a result, the position space $TE$ estimator no longer has minimum variance. We show however, in Appendix~\ref{sec:efficient_est} that the estimator normalization, variance and covariances change negligibly.

The single estimators are then combined to form the total CIP estimator
\bea
\hat{\Delta}_{LM}&=&\sum_{\alpha} w_L^{\alpha}\hat{\Delta}_{LM}^{\alpha},\;\;\; \label{eq:tot_estimator}
\eea
with inverse-covariance weights given by
\bea
w^{\alpha}_L=N_{L} \sum_{\beta}\left( \mathcal{M}_{L}^{-1}\right)^{\alpha,\beta},
\eea
where
\bea
N_{L}^{-1} \equiv{\sum_{\alpha\beta}\left(\mathcal{M}_{L}^{-1}\right)^{\alpha,\beta}} 
\label{eq:nldd_tot}
\eea
is the normalization that is required the make the total estimator unbiased. The variance of the total estimator $M_L^{\Delta\Delta}$ is the same as the normalization 
\beq
M_L^{\Delta\Delta} = N_L
\eeq
as long as we consistently use 
 \beq \label{eq:g_position_space}
	g_{L l l'}^{XZ}= 
\left\{
	\begin{array}{ll}
     	\bar{g}_{L l l'}^{XZ}, & \alpha = TE;\\
     	g_{L l l'}^{XZ, \rm mv}, 	& \rm other
	\end{array}
	\right.
\eeq
in the covariance matrix 
\bea
\mathcal{M}_{L}&&^{XZ, X'Z'}=N_{L}^{XZ}N_{L}^{X'Z'}\sum_{l l'}G_{l l'}g_{L l l'}^{XZ} \\ 
&&\left[ C_{l'}^{XX'}C_{l}^{ZZ'}g_{L l l'}^{X'Z'*}+\left(-1\right)^{l+l'+L}C_{l'}^{XZ'}C_{l}^{X'Z}g_{L l' l}^{X'Z'*} \right]. \notag\label{eq:covmat_total}
\eea
This result of the covariance matrix follows from Eq.~\ref{eq:single_est} where the CMB fields are taken to be built from Gaussian variates as described above.

Just like the single estimators, the total estimator is unbiased for Gaussian CMB realizations. In the absence of a true CIP signal, we expect $\langle \hat{\Delta}_{LM} \rangle = 0$ for the reconstructed maps. The power spectra however, have noise associated with the cosmic variance of the CMB modes. We study the noise distribution by first building the power spectrum estimators in each realization
\begin{equation}
\hat{M}_L^{XY} =  \frac{1}{2L+1} \sum_{M} \hat {X}_{LM}^{*} \hat{ Y}_{LM},
\label{eqn:noiserealization}
\end{equation} 
where $X,Y \in \{\Delta, T,E,B\}$.
Then we obtain the average $\langle \hat{M}_L^{XY}\rangle$ over 4000 realizations, verifying that the lensed spectra $C_L^{XY}$ are recovered for CMB fields $X,Y \in \{T,E,B\}$.

\begin{figure}
         \includegraphics[width=\columnwidth]{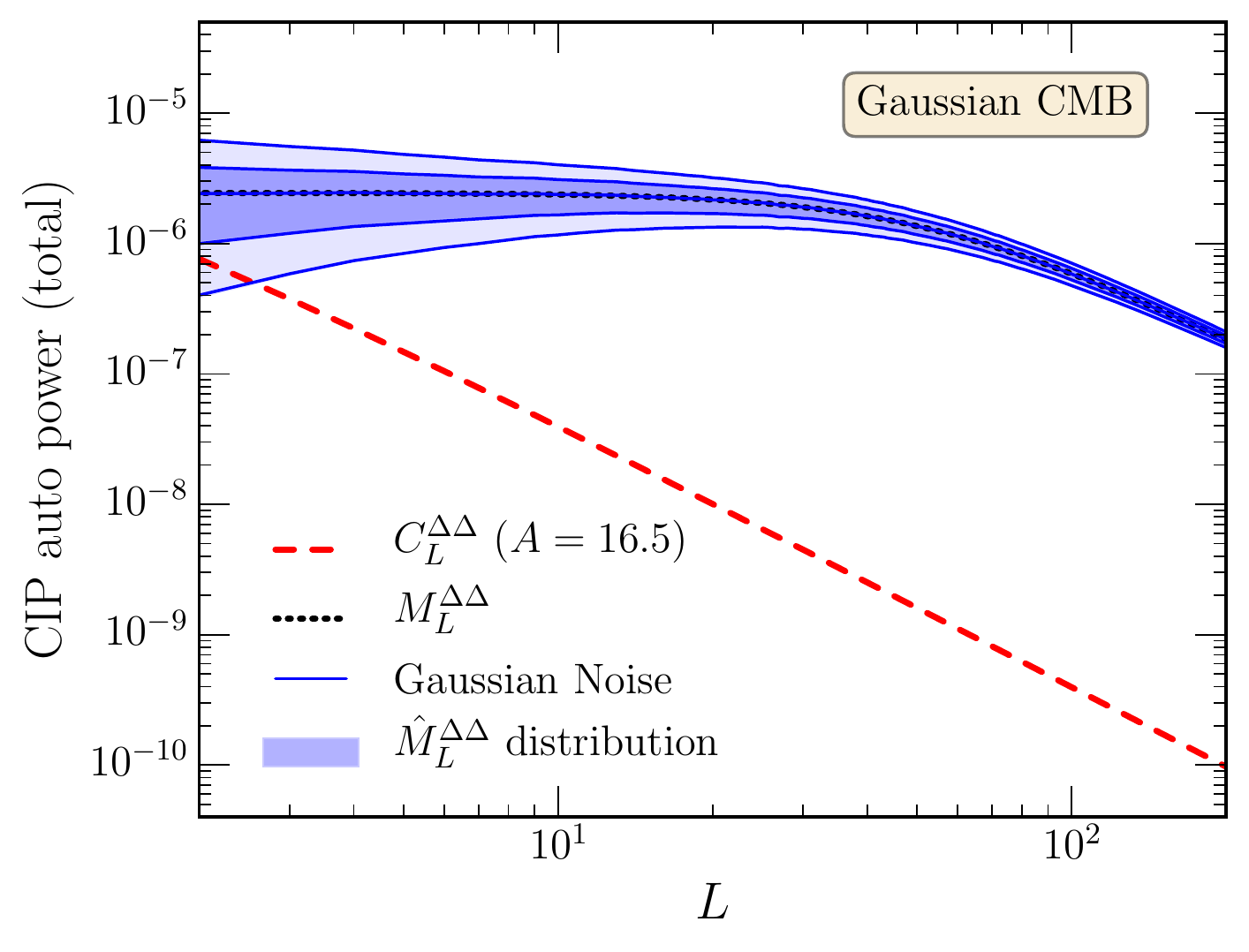}
          \caption[CIP noise auto power (Gaussian CMB)]{Total CIP estimator noise power $\hat{M}_{L}^{\Delta\Delta}$ for Gaussian CMB maps. 
          Shown are the mean (middle solid line), 68\% and 95\% confidence bands (shaded) of 4000 realizations of the total estimator in
         the absence of a CIP signal. The mean matches the theoretical expectation $M_L^{\Delta\Delta}$ (Eq.~\ref{eq:nldd_tot}). The confidence bands of $\hat{M}_{L}^{\Delta\Delta}$ match those from a $\chi^2$ distribution (solid lines) given the mean, expected for Gaussian estimator noise. For reference we show a true correlated CIP signal with $A=16.5$ (dashed line).
          }  \label{fig:DDg_tot}
\end{figure}

\begin{figure}
          \includegraphics[width=\columnwidth]{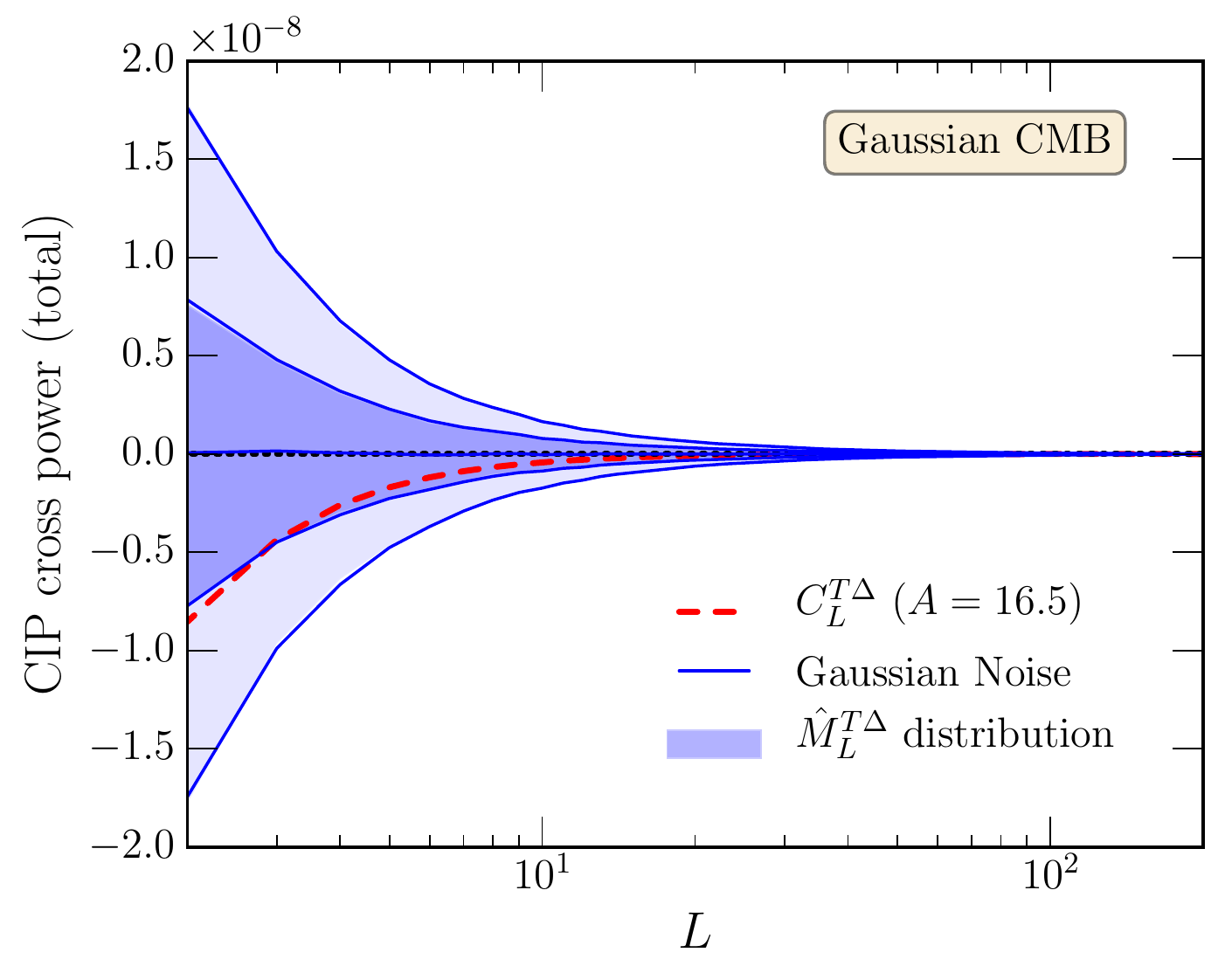}
          \caption{Total CIP estimator noise cross power $\hat{M}_{L}^{T\Delta}$ given Gaussian CMB maps.
          Shown are the mean (middle solid line), 68\% and 95\% confidence bands (shaded) of 4000 realizations of the estimator in
          the absence of a CIP signal.   The mean matches closely the expectation $M_L^{T\Delta} = 0$ (dotted line). The confidence bands of $\hat{M}_{L}^{T\Delta}$ match those from a Wishart distribution (solid lines) given the mean,  expected for Gaussian estimator noise.  For reference we show a true correlated CIP signal with $A=16.5$ (dashed line).
           }  \label{fig:TDg_tot}
\end{figure}

\begin{figure}
          \includegraphics[width=\columnwidth]{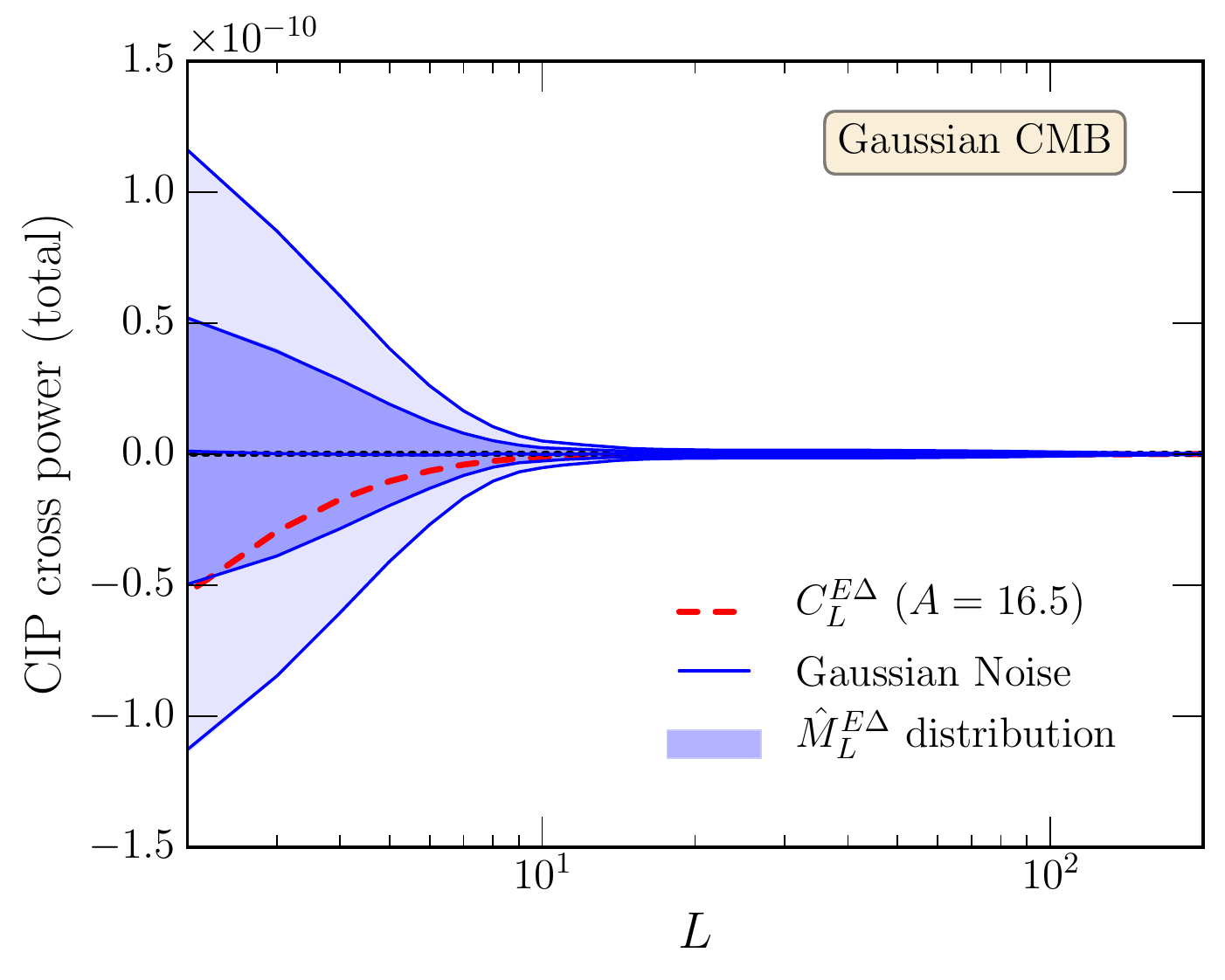}
          \caption{Total CIP estimator noise cross power $\hat{M}_{L}^{E\Delta}$ for Gaussian CMB maps.
          Shown are the mean (middle solid line), 68\% and 95\% confidence bands (shaded) of 4000 realizations of the estimator in
          the absence of a CIP signal.   The mean matches closely the expectation $M_L^{E\Delta} = 0$ (dotted line). The confidence bands of $\hat{M}_{L}^{E\Delta}$ match those of a Wishart distribution (solid lines) given the mean, expected for Gaussian estimator noise.  For reference we show a true correlated CIP signal with $A=16.5$ (dashed line).
          }  \label{fig:EDg_tot}
\end{figure}

For the CIP reconstruction, we plot the mean (middle blue line), 68\% and 95\% confidence bands (shaded bands) of the $\hat{M}_L^{\Delta\Delta}$, $\hat{M}_L^{T\Delta}$ and $\hat{M}_L^{E\Delta}$ distribution in Figs.~\ref{fig:DDg_tot}, \ref{fig:TDg_tot} and \ref{fig:EDg_tot} respectively. The mean agrees well with the ensemble average $M_L^{\Delta\Delta}$ of Eq.~\ref{eq:nldd_tot} and $M_L^{T\Delta} = M_L^{E\Delta} = 0$ (black dashed). In Fig.~\ref{fig:DDg_tot}, the total estimator noise power is dominated by white noise contributions from $TT, TE, EE$ at low-$L$ and by scale-invariant noise decreasing as $\sim L^{-2}$ of the $TB$, $EB$ estimators at high-$L$. Note that the addition of the $E$-mode polarization contributes to reducing the noise from $TT$ alone by about a factor of three. For the cross correlations, the improvement from adding polarization estimators is reflected in the relatively smaller width of the distribution.

At high-$L$, the relative scaling of $L^{-2}$ for the $B$ estimators are key to improving the total estimator noise. This scaling comes from the fact that the $B$ and non-$B$ estimators respond to CMB multipoles pairs $l, l'$ with odd and even $l+l'+L$ respectively. More specifically, the response function is proportional to 
\begin{eqnarray}
H_{l l'}^{L}& \propto
\begin{cases}
\sin (2\varphi_{ll'}), & l+l'+L~{\rm odd}, \\
\cos (2\varphi_{ll'}), &  l+l'+L~{\rm even} ,
\end{cases}
\end{eqnarray}
where $\varphi_{ll'}$ is the angle between the $l$ and $l'$ sides of the triangle, so in the squeezed limit $l,l' \gg L $ where the CIP signal dominates, $N_L^{XB} \sim (H_{l l'}^{L} )^{-2} \sim L^{-2}$. In Fig.~\ref{fig:plot_wl}, we plot the weights from the non-$B$ vs $B$ estimators in black and red respectively. For the Gaussian CMB considered here (solid lines), the total estimator becomes dominated by $B$ estimators for $L\gtrsim 50$.

The total CIP estimator is a linear combination of the single estimators, which are formed out of products of Gaussian variates. Although the individual product pairs are not Gaussian distributed, by the central limit theorem the linear combination of many such pairs tends to a Gaussian distribution given large enough numbers of pairs. To test the Gaussian approximation, we follow Ref.~\cite{Heinrich:2016gqe} to compute the expected $\chi^2$ and Wishart distributions for the auto and cross spectra respectively. In Figs.~\ref{fig:DDg_tot}, \ref{fig:TDg_tot} and~\ref{fig:EDg_tot}, we find that the confidence bands of the actual distribution (shaded) agree well with the Gaussian expectation (solid lines), indicating that the Gaussian noise is indeed a good approximation for the total estimator on Gaussian CMB maps. We have also verified that the same conclusion holds for the single estimators.

\begin{figure}
          \includegraphics[width=\columnwidth]{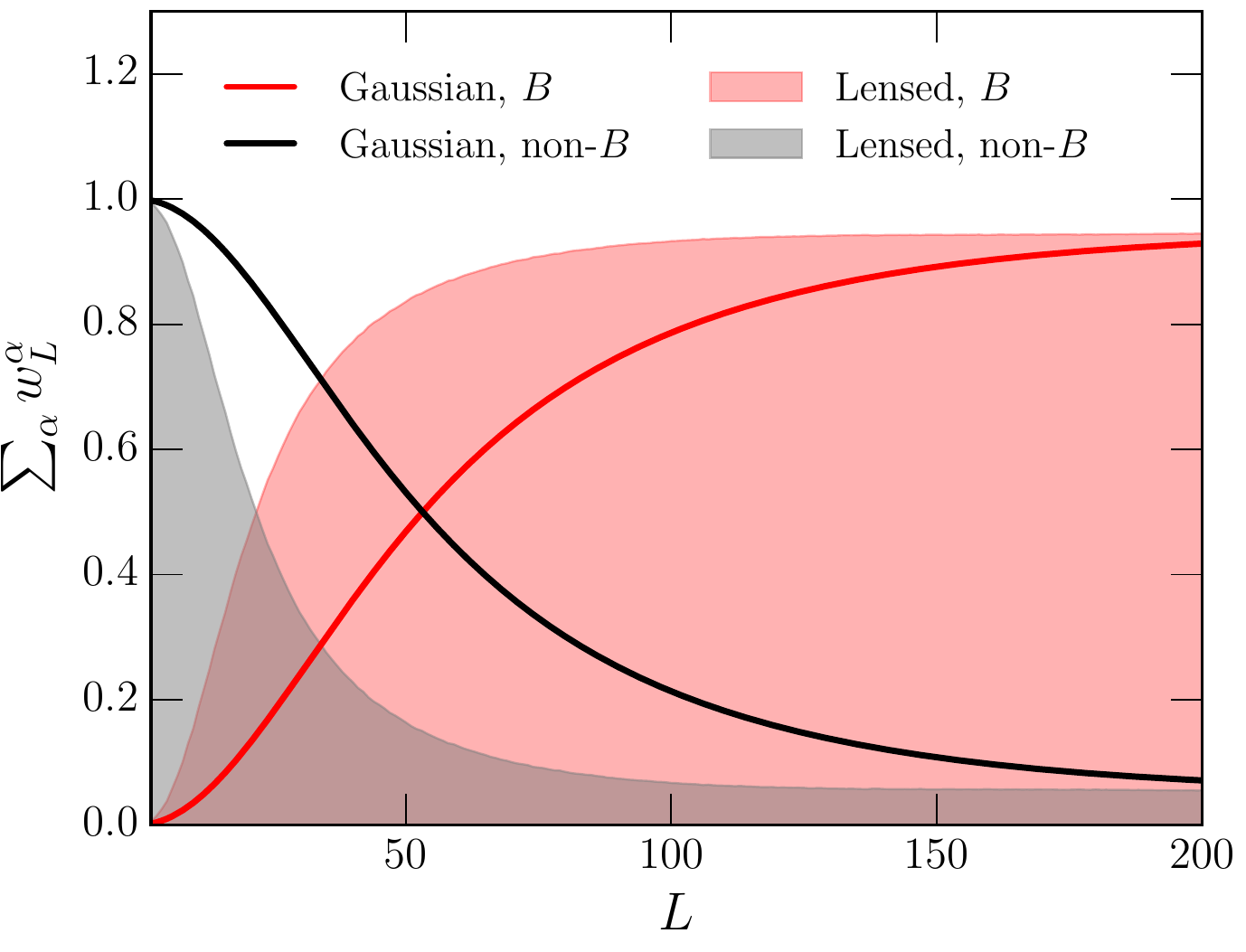}
          \caption{Inverse-covariance weights $\sum_{\alpha}w_L^{\alpha}$ for the $B$-based ($\alpha = TB, EB$, red) and non-$B$ based ($\alpha = TT, TE, EE$, black) estimators. For Gaussian CMB (solid lines), the total estimator is dominated by the $TB, EB$ at high-$L$ because of their nearly scale-invariant noise $\sim L^{-2}$; at low-L, the non-$B$ estimators dominate with their white noise. The crossing point happens at around $L \sim50$. For the properly lensed CMB (shaded), the non-Gaussian contributions from lensing to the estimator variance are more significant for the non-$B$ than for the $B$ estimators, so the former stops dominating at a small $L \lesssim 25$. 
         }
          \label{fig:plot_wl}
\end{figure}

\subsection{Lensing noise}

We now perform the same CIP reconstruction on a set of properly lensed CMB maps containing non-Gaussian lensing contributions, and study the resulting additional contribution to the estimator noise spectra.

To do so, we first simulate 4000 correlated realizations of the unlensed $\tilde{T}_{lm}$,
$\tilde{E}_{lm}$ and lensing potential $\phi_{lm}$ consistent with $C_l^{\tilde{T}\tilde{T}}$,
$C_l^{\tilde{E}\tilde{E}}$ and $C_l^{\phi\phi}$, and the cross-correlations $C_l^{\tilde{T}\tilde{E}}$,
$C_l^{\tilde{T}\phi}$ and $C_l^{\tilde{E}\phi}$ as supplied by CAMB using the method described
in Section~\ref{sec:gaussian}. Note that $\tilde{B}_{lm} = 0$ in the absence of tensor perturbations.

Using Lenspix, the pixel positions in the unlensed temperature maps and polarization tensor maps $\tilde {\mathcal{P}}_{ij}$ (formed from its $EB$ decomposition~\cite{Okamoto:2003zw}) are deflected according to the gradient of the lensing potential~\cite{Zaldarriaga:1998te,Hu:2001fa,Okamoto:2003zw}
\bea
 \hat T(\bn) = \tilde T(\bn+\nabla \phi),\\
\hat{\mathcal{P}}_{ij}(\bn) = \tilde {\mathcal{P}}_{ij}(\bn+\nabla \phi),
\eea
yielding the lensed maps $\hat{T}_{lm}$, $\hat{E}_{lm}$ and $\hat{B}_{lm}\neq0$.

Like CIPs in the SU approximation, the large-scale lenses also correlate the CMB anisotropies of different multipoles, albeit through a different mechanism remapping the angular positions of the CMB. As a result the CIP estimators pick up extra lensing signal and are no longer unbiased when averaged over CMB realizations with a fixed lensing potential
\begin{equation}
\langle \hat{\Delta}_{LM}^{\alpha}  \rangle\big|_\phi \ne 0, \;\;\;\; \langle \hat{\Delta}_{LM}  \rangle\big|_\phi \ne 0.
\end{equation}
Once averaged over random realizations of the lensing potentials we still recover $\langle \hat{\Delta}_{LM}^{\alpha}\rangle = \langle \hat{\Delta}_{LM}\rangle =  0$. The estimator power spectra, however, will retain the non-Gaussian lensing contributions through the connected part of the trispectrum
\begin{equation}
\langle \hat{M}_L^{\Delta^{\alpha}\Delta^{\alpha}}\rangle = \mathcal{M}_L^{\alpha, \alpha} + \mathcal{T}^{\alpha, \alpha}_{L}
\end{equation}

\begin{figure}
          \includegraphics[width=\columnwidth]{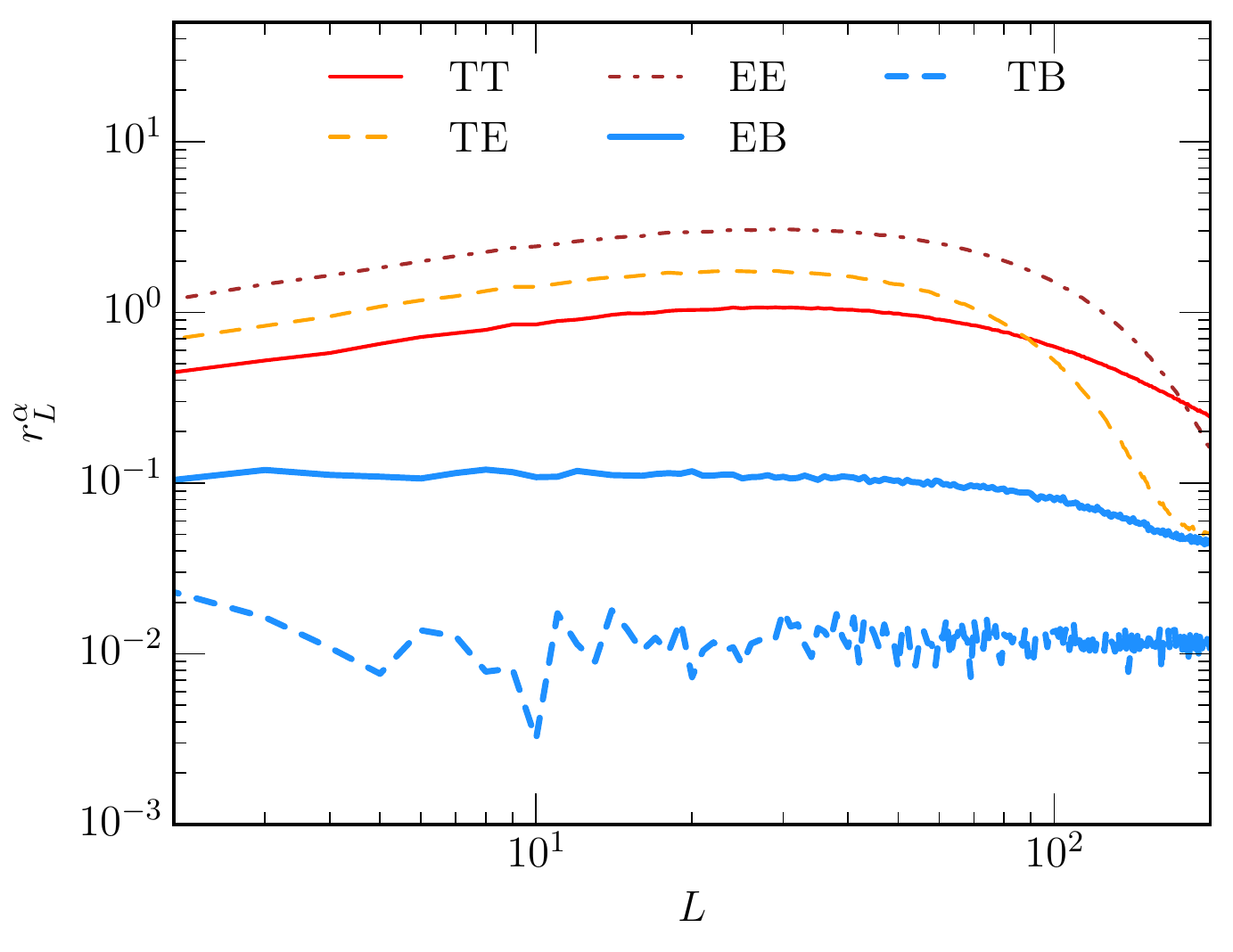}
          \caption{Absolute ratio $r_L^{\alpha}$ of non-Gaussian lensing to Gaussian CMB contributions to the noise power spectrum of single estimators. The $TB$, $EB$ estimators (lower lines) are much less contaminated (only at 1\% and 10\% level respectively) by the non-Gaussian contributions of lensing. For the non-$B$ estimators (upper lines), the lensing and Gaussian CMB noises are of the same order for $L\lesssim100$. 
           }  \label{fig:DD_bias_single}
\end{figure}

In Fig.~\ref{fig:DD_bias_single}, we plot in absolute ratio of non-Gaussian lensing contributions to those expected from Gaussian CMB for the noise power of single estimators
\begin{equation}
r_L^{\alpha} = \frac{\mathcal{T}^{\alpha, \alpha}_{L}}{\mathcal{M}_L^{\alpha, \alpha}}
\end{equation}
The ratio is roughly flat for each single estimator on scales relevant for the SU limit $L \lesssim 100$, meaning that the lensing induced noise has a similar spectrum shape to the Gaussian CMB contributions. Just like $TT$, the lensing contamination in $TE$ and $EE$ are about the same level as the Gaussian CMB part. In contrast, the lensing noise in $TB$ and $EB$ are only the percent level and 10\% level respectively of the Gaussian CMB contributions. 

Given that $TB$ and $EB$ have significantly less lensing noise, it is desirable to weigh the single estimators accordingly to lensing-included covariance derived from simulations in lieu of Eq.~\ref{eq:covmat_total}. These weights are shown as shaded regions in Fig.~\ref{fig:plot_wl}. We see that the $B$-estimators now have slightly higher weight at low-$L$, and start dominating the total at a smaller $L \sim 25$. 

\begin{figure}
          \includegraphics[width=\columnwidth]{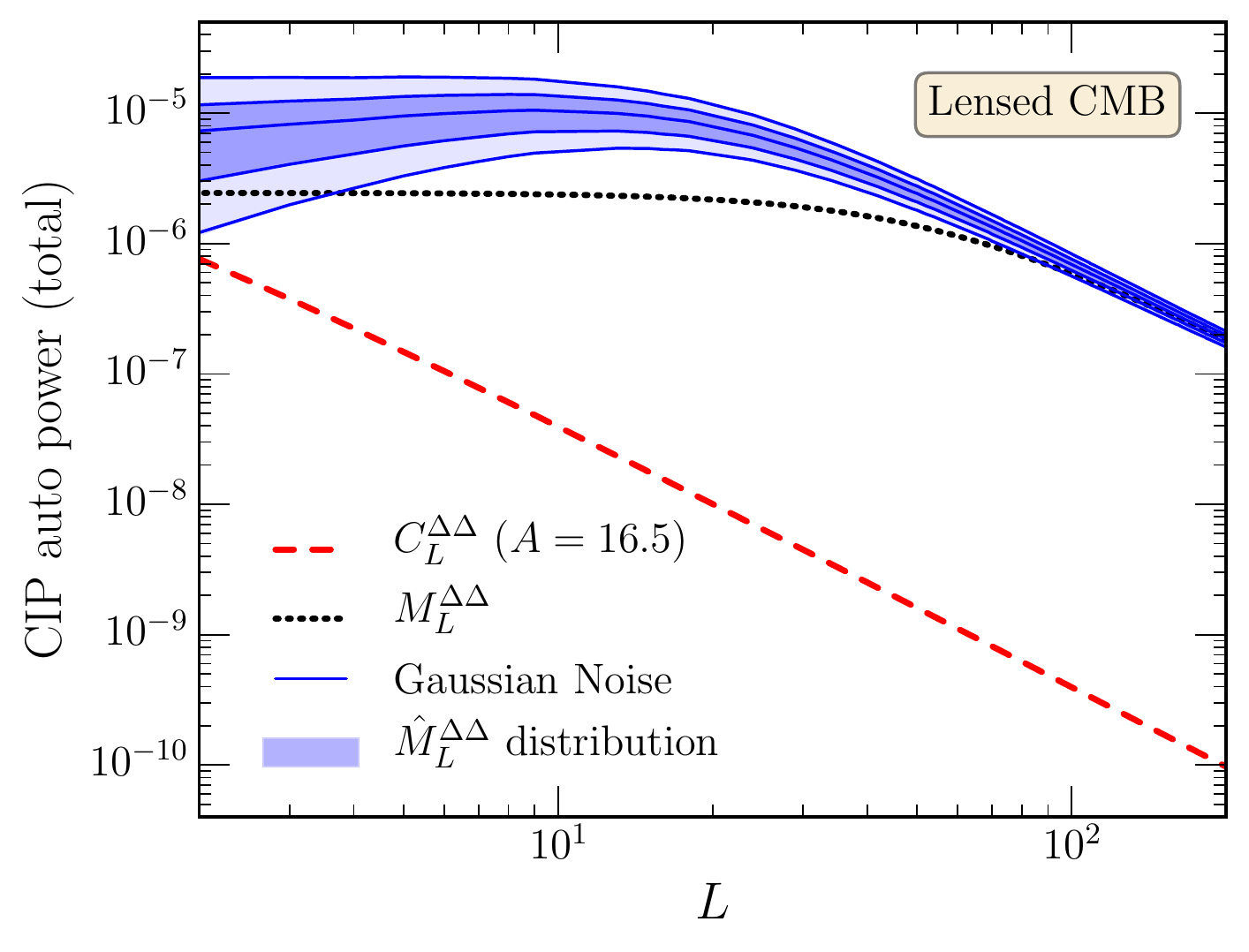}
          \caption{Total CIP estimator noise power $\hat{M}_{L}^{\Delta\Delta}$ for lensed CMB maps. 
          Shown are the mean (middle solid line), 68\% and 95\% confidence bands (shaded) of 4000 realizations of the total estimator in
          the absence of a CIP signal. On large scales $L\lesssim40$, the mean is about three times larger with lensing effects than without (dotted line). The lensing bias is smaller for smaller scales because there the total estimator starts to be dominated by the less biased $B$ estimators. Despite the non-Gaussian contributions of lensing to CMB fields, the distribution of the CIP estimator noise power match closely the $\chi^2$ expectation for Gaussian estimator noise (solid lines) even out to the 95\% tail. For reference we show a true correlated CIP signal with $A=16.5$ (dashed line).
          }  \label{fig:DDng_tot_sim}
\end{figure}

\begin{figure}
          \includegraphics[width=\columnwidth]{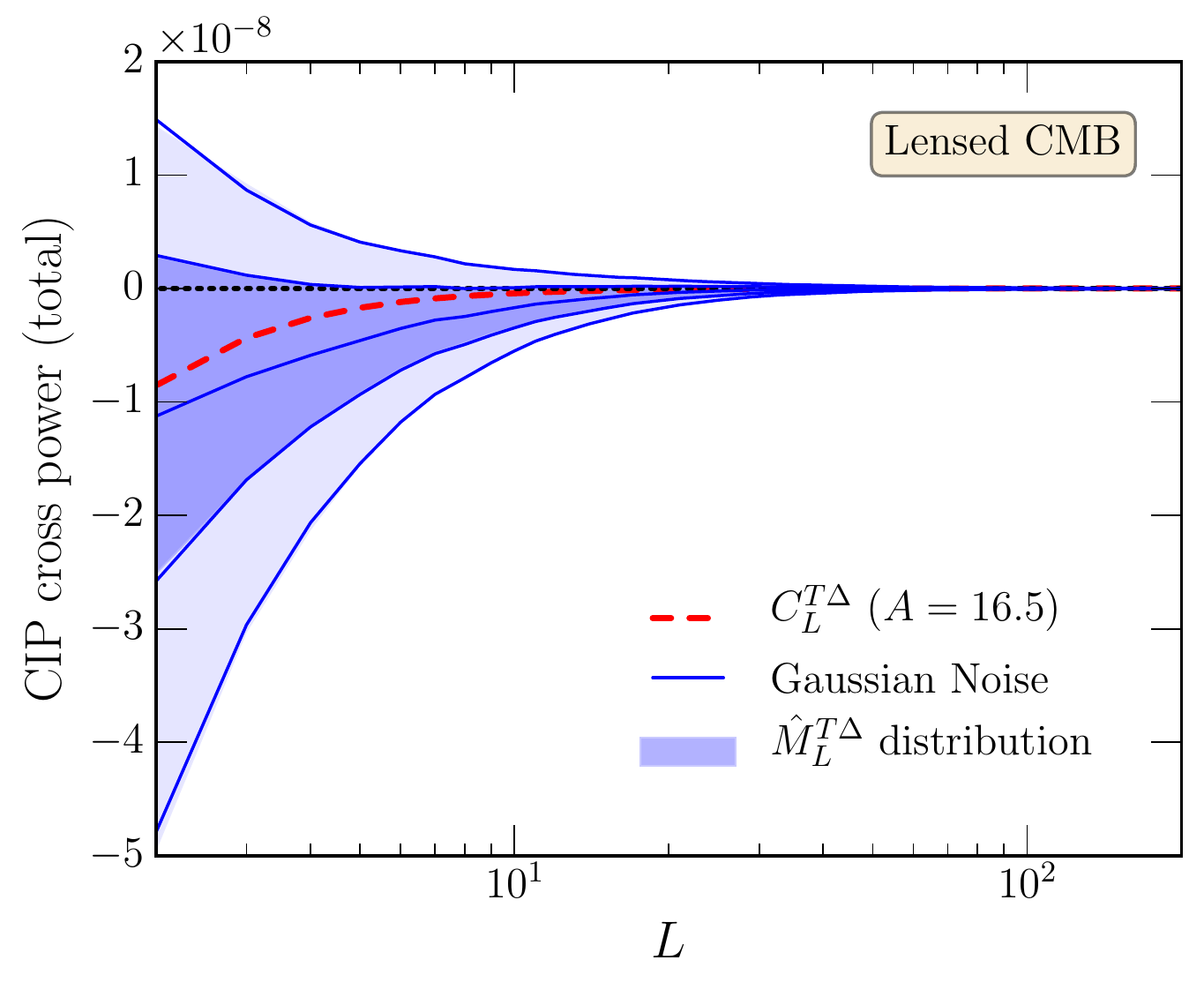}
          \caption{Total CIP estimator noise cross power $\hat{M}_{L}^{T\Delta}$ for lensed CMB maps.
          Shown are the mean (middle solid line), 68\% and 95\% confidence bands (shaded) of 4000 realizations of the total estimator in
          the absence of a CIP signal.  The contamination to the zero expectation of Gaussian CMB (dotted line) comes from the lensing-ISW correlation~\cite{Smith:2006ud,Lewis:2011fk,Kim:2013nea,Ade:2015dva}. The confidence bands again match the expectations for Gaussian estimator noise (solid lines), i.e. those of a Wishart distribution for the cross power given the mean.  For reference we show a true correlated CIP signal with $A=16.5$ (dashed line).
           }  \label{fig:TDng_tot_sim}
\end{figure}

\begin{figure}
         \includegraphics[width=\columnwidth]{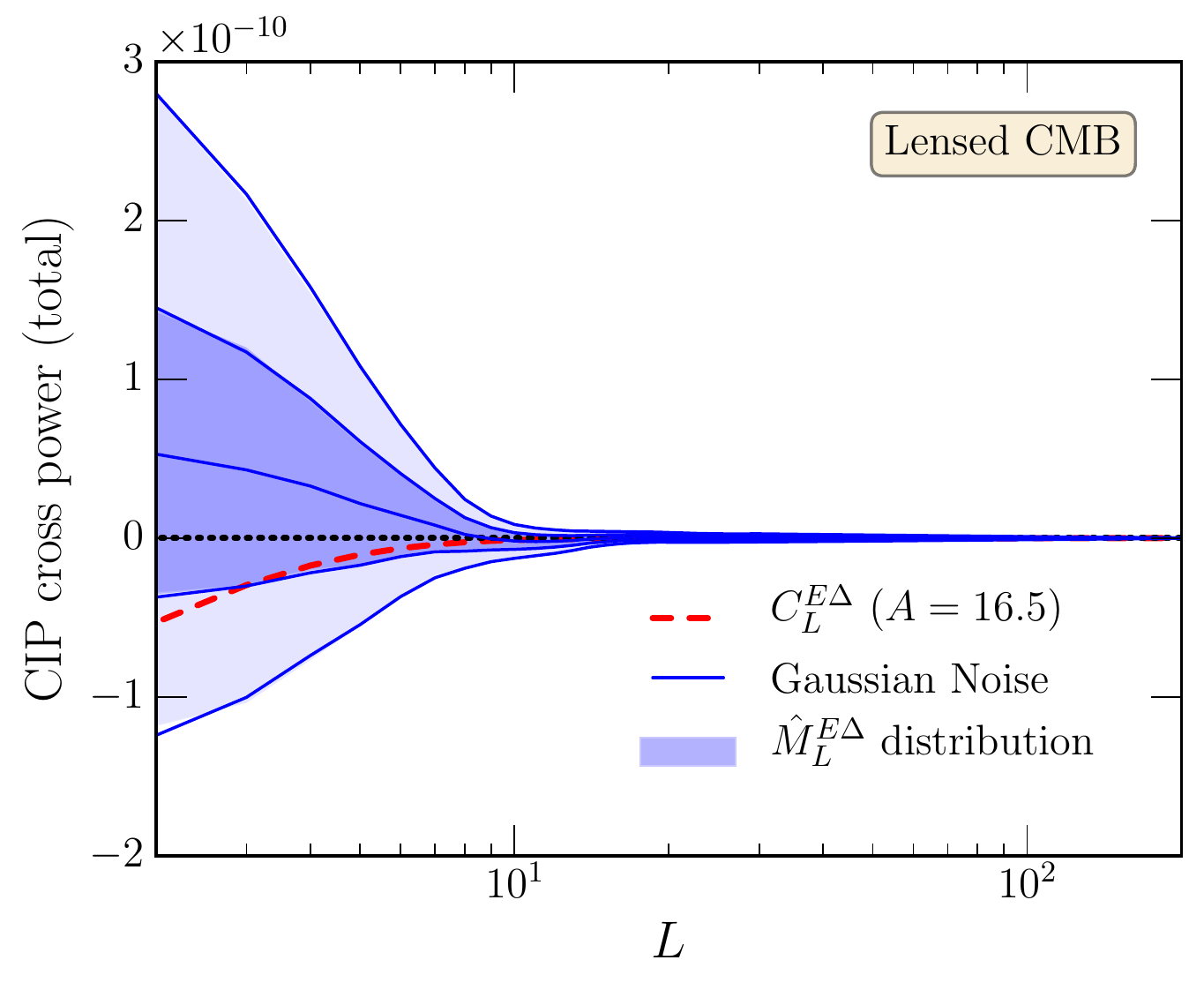}
          \caption{Total CIP estimator noise cross power $\hat{M}_{L}^{E\Delta}$ for lensed CMB maps.
          Shown are the mean (middle solid line), 68\% and 95\% confidence bands (shaded) of 4000 realizations of the estimator in
          the absence of a CIP signal.   The contamination to the zero expectation from Gaussian CMB (dotted line) is dominated by large-scale correlation between $E$-polarization and the lensing potential~\cite{Lewis:2011fk}.  The confidence bands again match the expectations for Gaussian estimator noise (solid lines), i.e. those of a Wishart distribution for the cross power given the mean.   For reference we show a true correlated CIP signal with $A=16.5$ (dashed line).
           }  \label{fig:EDng_tot_sim}
\end{figure}

Using these weights, we form the total estimator and plot the mean (middle blue line), 68\% and 95\% confidence bands (shaded bands) of its noise power $\hat{M}_L^{\Delta\Delta}$, $\hat{M}_L^{T\Delta}$ and $\hat{M}_L^{E\Delta}$ in Figs~\ref{fig:DDng_tot_sim},~\ref{fig:TDng_tot_sim} and~\ref{fig:EDng_tot_sim} respectively.  At $L \lesssim 40$, the total noise power is about three times larger with non-Gaussian lensing contributions than without. Beyond this range, the lensing contributions to the mean becomes comparable or smaller than the Gaussian CMB contributions as the $B$-estimators dominate the weight. Note that for $L \gtrsim100$ the bias reduces significantly, but this is also beyond the SU limit where little CIP signal exists. 

Compared to the zero expectation for Gaussian CMB, the cross-spectra $\hat{M}_L^{T\Delta}$ here acquires a lensing-ISW~\cite{Smith:2006ud,Lewis:2011fk,Kim:2013nea,Ade:2015dva} contamination on large scales. This is because in the absence of a CIP signal, the estimator is basically reconstructing a lensing signal. Similarly, $\hat{M}_L^{E\Delta}$ mean now oscillates with a similar shape to $C_L^{E\phi}$, which is dominated by the correlation between large-scale reionization signal in $E$ and low-$z$ matter density fluctuations contributing to the lensing potential~\cite{Lewis:2011fk}. 

Even with the non-Gaussian lensing contributions, the approximation that the total estimator noise is Gaussian still holds. We find good agreement between the 68\% and 95\% confidence bands of the distribution and the Gaussian noise expectations (solid lines). We have also verified that the same is true for the single estimators.

To further test the Gaussian noise properties, we verify that the covariance of the noise power have negligible off-diagonal correlations. We start by building the covariance matrix 
\beq
C_{ij} \equiv \langle \hat{M}_{L_{i}}^{\Delta\Delta} \hat{M}_{L_{j}}^{\Delta\Delta}\rangle - \langle \hat{M}_{L_{i}}^{\Delta\Delta} \rangle \langle \hat{M}_{L_{j}}^{\Delta\Delta}\rangle.
\eeq
We plot the correlation matrix
\beq
R_{ij} = \frac{C_{ij}}{\sqrt{C_{ii}C_{jj}}}
\eeq
in Fig.~\ref{fig:R} where the off-diagonal elements in the range $L\in[2,200]$ do not exceed 0.065. They fluctuate around a negligible mean of $R = 3.5 \times 10^{-4}$ with a root-mean-square (r.m.s.) of $\sigma_R = 0.016$. The scaling of the r.m.s. is consistent with what is expected from a finite size of realizations, i.e.  $\sigma_R \sim n_{\rm sim}^{1/2}$ as shown in Fig.~\ref{fig:sigma_R}.

\begin{figure}
          \includegraphics[width=\columnwidth]{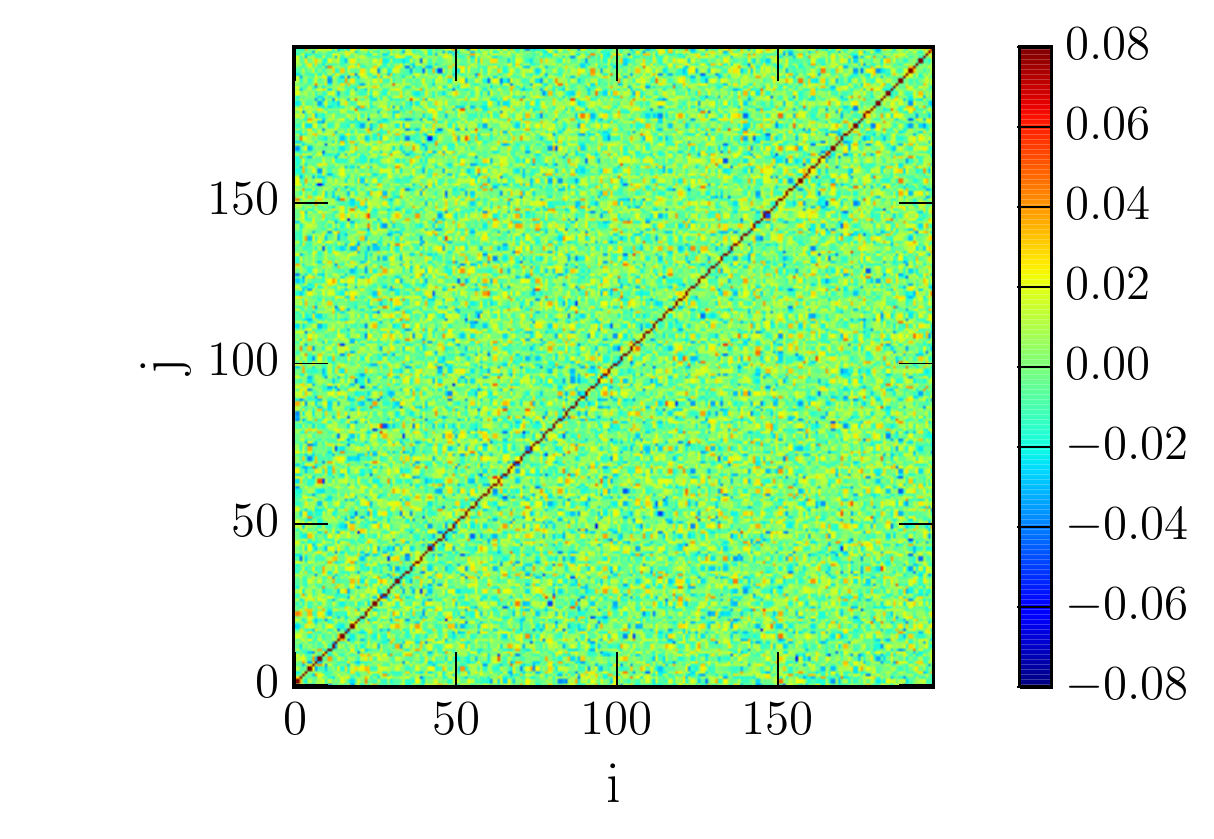}
          \caption[Correlation matrix between CIP noise power of different multipoles]{Correlation matrix $R_{ij}$ between different multipoles $L_i$ and $L_j$ of the total CIP noise power $\hat{M}_{L}^{\Delta\Delta}$ using 4000 realizations of the lensed CMB maps. The off-diagonal correlations have a negligible mean
         of $R = 3.5\times10^{-4}$ with an r.m.s. fluctuation~$\sigma_R=0.016$ consistent with
         the finite sample size.
           }  \label{fig:R}
\end{figure}

\begin{figure}
          \includegraphics[width=\columnwidth]{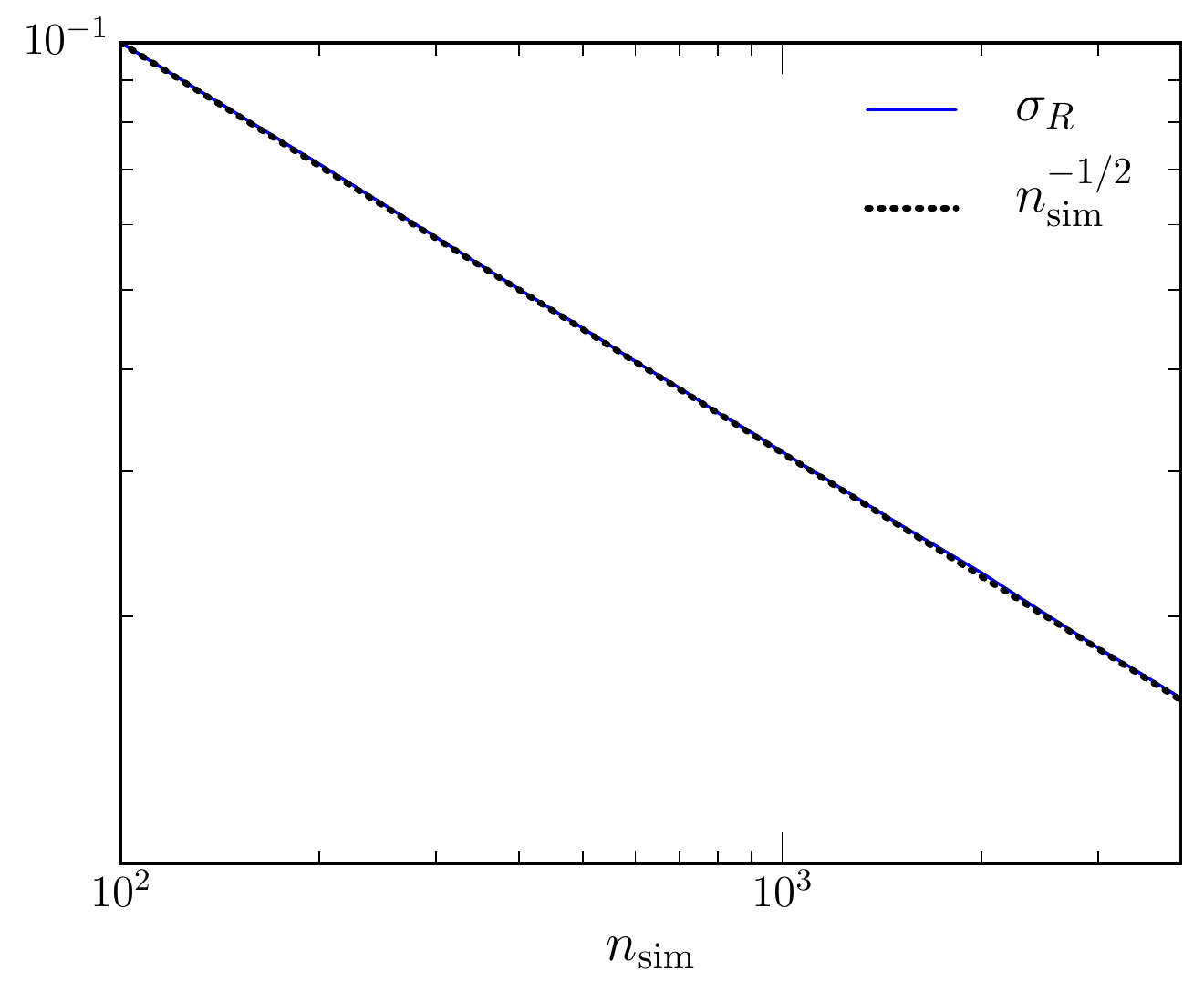}
          \caption[R.M.S. fluctuations of off-diagonal elements in the correlation matrix]{The r.m.s fluctuations $\sigma_R$ (solid) in the off-diagonal correlations $R_{ij}$ of CIP estimator noise power vs~the number of simulations $n_{\mathrm{sim}}$ for multipoles $L \in [2,200]$. The scaling of $\sigma_R$ agrees well with that expected from finite sampling $n_{\mathrm{sim}}^{-1/2}$ (dotted), again showing no hints of significant deviations from Gaussian estimator noise.
                   }  \label{fig:sigma_R}
\end{figure}

\section{Forecasts} \label{sec:forecasts}

We have seen previously that for CVL measurements of CMB temperature and polarization out to
$l_{\rm max} = 2500$, the noise power of the total CIP estimator noise is nearly three times
larger with than without lensing contributions for at least up to $L\sim40$.
In this section, we evaluate the impact of this additional lensing noise on CIP detectability by means of Fisher matrix techniques.

We have shown that the CIP estimator noise, even with non-Gaussian effects from lensing, can still be treated as nearly Gaussian distributed and with no correlation between different multipoles of the noise power. Under these approximations, we construct the Fisher matrix with a single entry to evaluate the error $\sigma_A$ in the CIP correlation amplitude $A$ from the observed CIP power spectra 
\beq 
	\sigma_A^{-2}  = \sum_{L=2}^{L_{\rm max}}  \sum_{X\Delta,X'\Delta} 
	\frac{\partial C_L^{X\Delta}}{\partial A}	\left({\bf C}^{-1}_{L}
	\right)_{X\Delta,X'\Delta} \frac{\partial C_L^{X'\Delta}}{\partial A},
	\label{eq:fisher}
\eeq
where $X,X' \in  \{\Delta, T, E\}$ and ${\bf C}_L$ is the covariance matrix 
\beq
		{\bf C}^{X\Delta,X'\Delta}_{L} =\frac{{ \tilde{C}_L^{XX'} \tilde{C}_{L}^{\Delta\Delta} + \tilde{C}_L^{X\Delta} \tilde{C}_{L}^{X'\Delta}}}{2L+1} .
\eeq
The covariance here includes both the CIP sample variance and the reconstruction noise from Gaussian or lensed CMB simulations of section~\ref{sec:simulations}, i.e.
\bea
&&\tilde{C}_L^{\Delta\Delta} = C_L^{\Delta\Delta} + \langle \hat{M}_L^{\Delta\Delta} \rangle, \\
&&\tilde{C}_L^{X\Delta} = C_L^{X\Delta} + \langle \hat{M}_L^{X\Delta} \rangle,\\
&&\tilde{C}_L^{XX'} = C_L^{XX'}, 
\eea
where $X, X'\in\{T,E\}$. Note that $\sigma_A$ depends on the strength of the signal through the CIP sample variance, so we evaluate the detection threshold at $A = 2\sigma_A$.

Taking $L_{\rm max} = 100$ as appropriate for the separate-universe approximation,
we obtain 2$\sigma_A = 12.2$ for the total estimator with CVL measurements of temperature and polarizations.
This is a factor of 1.5 higher than the 2$\sigma_A = 8.3$ threshold if lensing was not accounted for. 
For comparison, a less optimal weighting for the total estimator with Eq.~\ref{eq:covmat_total} would have given a threshold that is 1.8 times higher. Finally, with lensing noise accounted for in the total CIP estimator, the $4\sigma$ projection for the maximal CIP $A \approx 16.5$ scenario of the curvaton model reduces to $2.7\sigma$ for a cosmic-variance-limited experiment.

Taking $L_{\rm{max}} = 200$, we find that a smaller degradation with lensing, a factor of 1.2
from $2\sigma_A = 6.2$ to 7.5, due to decreasing lensing bias after $L_{\rm max} \sim 100$.
In addition, because precisely measuring the large-angle E-modes could be difficult with ground-based
experiments, we evaluate the detection threshold dropping all the correlations at $L<30$.
With $L_{\rm min}=30$, we find that the maximal CIP case would still be detected at $2.3\sigma$ with $2\sigma_A =14.4$ for
the CVL experiment.

\section{Conclusion} \label{conclusion}

In this paper, we evaluated for the first time the lensing bias to measurements of CIPs using CMB polarization and quantified the impact of lensing on CIP detectability for a cosmic-variance-limited experiment. 

We found that the polarization-included total CIP estimator has a noise power that is about three times larger with than without lensing contamination on $L\lesssim40$. In the cross-correlations of CIPs with temperature and $E$ mode polarization, lensing contamination follows the shape of ISW-lensing and reionization-lensing correlations on large scales. In addition, we found that $TB$ and $EB$ estimators are much less biased by lensing (only at the 1\% and 10\% level respectively in the auto noise power), even though they have larger noise from the cosmic variance of the CMB modes alone. So measuring the cross-spectra with $B$-estimators individually could provide a consistency test for determining the sign of correlated CIPs.

Although the lensing contributions to the CMB fields are non-Gaussian, we showed that their effect on the quadratic
estimators is to good approximation Gaussian noise in the total as well as the single estimators.
We further tested the Gaussian noise assumption by showing that the different multipoles
of the noise power are negligibly correlated at~$<0.065$ over the range $L \in [2 ,200]$.
The off-diagonal elements in the correlation matrix fluctuate around a mean of $R = 3.5\times10^{-4}$
with r.m.s consistent with the finite size of simulations.

While the use of polarization dramatically increases the CIP detectability compared to temperature only
measurements, there is still a relative degradation for polarization measurements once lensing noise is
included. Treating the estimator noise as Gaussian independent noise for each multipole of the noise power,
we found that the detection threshold of a CVL experiment increased a factor of 1.5 from $2\sigma_A = 8.3$ to 12.2 because of lensing, corresponding to $2.7\sigma$ detection for the maximal CIP $A\approx16.5$ scenario of the curvaton model. Taking $L_{\rm{min}} = 30$ gives $2\sigma_A =14.4$ which is still a $2.3\sigma$ for the $A\approx16.5$ scenario. Here we have used CVL measurements of temperature and polarization out to $l = 2500$ and fixed all other cosmological parameters. 

The next step in assessing the CIP detectability for a realistic CMB experiment would be to simulate the lensing bias dependence on instrument noise and sky masks. For a nearly CVL experiment like CMB Stage-4, one might expect a similar factor of degradation to the CVL experiment, bringing down the 3$\sigma$ projection to about 2$\sigma$ for the largest CIP signal $A \approx 16.5$ in the curvaton model. In addition, while we conservatively considered CMB multipoles up to $l = 2500$, future CMB measurements of $E$-mode polarization have the potential of reaching out to $l = 4000$. It would be interesting to study the impact on quadratic estimators with non-uniform $l_{\rm max}$ for $T$, $E$ and $B$ observations as well as its implications for lensing contamination.

An alternative route for removing lensing contamination to CIPs may be to use delensed
CMB maps~\cite{Larsen:2016wpa,Carron:2017vfg,Manzotti:2017net}. One concern while using
internally delensed maps with a lensing template reconstructed from the CMB itself is the
partial removal of the CIP signal during the delensing process. As the $B$-estimators for lensing would also be the least contaminated by CIPs, one may use  the $EB$ estimator with optimized weights to construct the lensing template. The prospect of the delensing method for CMB measurements of CIPs still remains to be evaluated in comparison to the debiasing method presented in this paper.

 \begin{acknowledgments}
I would like to thank Wayne Hu for instructive discussions, Daniel Grin for
kindly providing the derivative power spectra used in this work, Christopher Heinrich,
Daniel Holz, Liantao Wang and Abigail Vieregg for useful discussions and feedback on the manuscript. 
C.H. was supported by NASA ATP NNX15AK22G, U.S. Dept. of Energy Contract No. DE-FG02-13ER41958, and the
Kavli Institute for Cosmological Physics at the University of Chicago through Grants No. NSF PHY-0114422 and No. NSF PHY-0551142.
Computing resources were provided by the University of Chicago Research Computing Center.
\end{acknowledgments}

\appendix

\section{Efficient CIP Estimator in Position Space} \label{sec:efficient_est}

The harmonic-space forms for the CIP estimators are computational costly $\mathcal{O}(l_{\rm max}^3)$. For CIP reconstruction in this paper, we adopt the more efficient $\mathcal{O}(l_{\rm max}^2)$ position-space forms listed below:

\beq
\hD\LM\TT = N_L\TT \int d\hn \, Y\cc\LM (\hn) \,_0 A_{TT} \, _0 A_{TT}^{TdT}, 
\eeq

\beq
\hD\LM\EE = \frac{1}{2} N_L\EE \int d\hn \, Y\cc\LM (\hn) \,( _{+2} A_{EE} \, _{-2} A_{EE}^{EdE} + \mathrm{c.c.}),
\eeq

\bea
\hD\LM\TE &=&N_L\TE \int d\hn \, Y\cc\LM (\hn) \,  \notag  \\
&&( \frac{1}{2}\left[( _{+2} A_{EE} \, _{-2} A_{TT}^{TdE} - _{+2} A_{ET} \, _{-2} A_{ET}^{TdE}) +\mathrm{c.c.}  \right] \notag \\
&&+  \left[ \,_{0} A_{TT} \, _{0} A_{EE}^{EdT} - \,_{0} A_{ET} \, _{0} A_{TE}^{TdE}) \right] ),
\eea

\beq
\hD\LM\TB = \frac{1}{2} N_L\TB \int d\hn \, Y\cc\LM (\hn) \,(i _{+2} A_{BB} \, _{-2} A_{TT}^{TdE} + \mathrm{c.c.}), 
\eeq

\beq
\hD\LM\EB =\frac{1}{2} N_L\EB \int d\hn \, Y\cc\LM (\hn) \,(i _{+2} A_{BB} \, _{-2} A_{EE}^{EdE} + \mathrm{c.c.}), 
\eeq

where 

\beq
_{\pm s} A_{XX'} = \sum_{lm} \frac{C_l^{XX'}}{C_l^{XX}C_l^{X'X'}}  X\lm \, _{\pm s} Y\lm,
\eeq
\beq
_{\pm s} A_{XX'}^{YdZ} =\sum_{lm} \frac{C_l^{XX'}}{C_l^{XX}C_l^{X'X'}}  C_l^{Y,dZ} X\lm \, _{\pm s} Y\lm, 
\eeq

\beq
\left[N_{L}^{XZ}\right]^{-1}=\sum_{l l'}G_{l l'}S_{l l'}^{L,XZ}g_{L l l'}^{ XZ, \rm}, \label{eq:single_norm_position_space}
\eeq

and

 \beq \label{eq:g_position_space}
	g_{L l l'}^{XZ}= 
\left\{
	\begin{array}{ll}
     	\bar{g}_{L l l'}^{XZ}, & \alpha = TE;\\
     	g_{L l l'}^{XZ, \rm mv}, 	& \rm other.
	\end{array}
	\right.
\eeq

These expressions are mathematically equivalent to the harmonic-space forms except for the $TE$ estimator, which cannot be written as a product of maps unless we drop the second term in the denominator of $g_{Lll'}^{TE, \rm mv}$
\bea
g_{Lll'}^{TE, \rm mv} &\propto& \frac{1}{C_{l'}\TT C_l\EE C_{l}\TT C_{l'}\EE - (C_l\TE C_{l'}\TE)^2}  \notag \\ 
\rightarrow \bar{g}_{Lll' }^{TE} & \propto & \frac{1}{C_{l'}\TT C_l\EE C_{l}\TT C_{l'}\EE}.
\eea

This approximation leads to only percent level differences in the estimator normalization and its covariance with other estimators ($<1.1\%$ and $<0.3\%$ respectively), and to a vanishing fractional difference $<0.03\%$ in the $TE$ estimator variance. 

\bibliography{chen_spires}

\begin{thebibliography}{42}%
\makeatletter
\providecommand \@ifxundefined [1]{%
 \@ifx{#1\undefined}
}%
\providecommand \@ifnum [1]{%
 \ifnum #1\expandafter \@firstoftwo
 \else \expandafter \@secondoftwo
 \fi
}%
\providecommand \@ifx [1]{%
 \ifx #1\expandafter \@firstoftwo
 \else \expandafter \@secondoftwo
 \fi
}%
\providecommand \natexlab [1]{#1}%
\providecommand \enquote  [1]{``#1''}%
\providecommand \bibnamefont  [1]{#1}%
\providecommand \bibfnamefont [1]{#1}%
\providecommand \citenamefont [1]{#1}%
\providecommand \href@noop [0]{\@secondoftwo}%
\providecommand \href [0]{\begingroup \@sanitize@url \@href}%
\providecommand \@href[1]{\@@startlink{#1}\@@href}%
\providecommand \@@href[1]{\endgroup#1\@@endlink}%
\providecommand \@sanitize@url [0]{\catcode `\\12\catcode `\$12\catcode
  `\&12\catcode `\#12\catcode `\^12\catcode `\_12\catcode `\%12\relax}%
\providecommand \@@startlink[1]{}%
\providecommand \@@endlink[0]{}%
\providecommand \url  [0]{\begingroup\@sanitize@url \@url }%
\providecommand \@url [1]{\endgroup\@href {#1}{\urlprefix }}%
\providecommand \urlprefix  [0]{URL }%
\providecommand \Eprint [0]{\href }%
\providecommand \doibase [0]{http://dx.doi.org/}%
\providecommand \selectlanguage [0]{\@gobble}%
\providecommand \bibinfo  [0]{\@secondoftwo}%
\providecommand \bibfield  [0]{\@secondoftwo}%
\providecommand \translation [1]{[#1]}%
\providecommand \BibitemOpen [0]{}%
\providecommand \bibitemStop [0]{}%
\providecommand \bibitemNoStop [0]{.\EOS\space}%
\providecommand \EOS [0]{\spacefactor3000\relax}%
\providecommand \BibitemShut  [1]{\csname bibitem#1\endcsname}%
\let\auto@bib@innerbib\@empty
\bibitem [{\citenamefont {Ade}\ \emph {et~al.}(2013)\citenamefont {Ade} \emph
  {et~al.}}]{Ade:2013uln}%
  \BibitemOpen
  \bibfield  {author} {\bibinfo {author} {\bibfnamefont {P.}~\bibnamefont
  {Ade}} \emph {et~al.} (\bibinfo {collaboration} {Planck Collaboration}),\
  }\href@noop {} {\  (\bibinfo {year} {2013})},\ \Eprint
  {http://arxiv.org/abs/arXiv:1303.5082} {arXiv:1303.5082} \BibitemShut
  {NoStop}%
\bibitem [{\citenamefont {Ade}\ \emph {et~al.}(2015{\natexlab{a}})\citenamefont
  {Ade} \emph {et~al.}}]{Ade:2015lrj}%
  \BibitemOpen
  \bibfield  {author} {\bibinfo {author} {\bibfnamefont {P.~A.~R.}\
  \bibnamefont {Ade}} \emph {et~al.} (\bibinfo {collaboration} {Planck}),\
  }\href@noop {} {\  (\bibinfo {year} {2015}{\natexlab{a}})},\ \Eprint
  {http://arxiv.org/abs/1502.02114} {arXiv:1502.02114} \BibitemShut {NoStop}%
\bibitem [{\citenamefont {Ade}\ \emph {et~al.}(2015{\natexlab{b}})\citenamefont
  {Ade} \emph {et~al.}}]{Ade:2015ava}%
  \BibitemOpen
  \bibfield  {author} {\bibinfo {author} {\bibfnamefont {P.~A.~R.}\
  \bibnamefont {Ade}} \emph {et~al.} (\bibinfo {collaboration} {Planck}),\
  }\href@noop {} {\  (\bibinfo {year} {2015}{\natexlab{b}})},\ \Eprint
  {http://arxiv.org/abs/1502.01592} {arXiv:1502.01592} \BibitemShut {NoStop}%
\bibitem [{\citenamefont {Bond}\ and\ \citenamefont
  {Efstathiou}(1984)}]{Bond:1984fp}%
  \BibitemOpen
  \bibfield  {author} {\bibinfo {author} {\bibfnamefont {J.~R.}\ \bibnamefont
  {Bond}}\ and\ \bibinfo {author} {\bibfnamefont {G.}~\bibnamefont
  {Efstathiou}},\ }\href {\doibase 10.1086/184362} {\bibfield  {journal}
  {\bibinfo  {journal} {Astrophys. J.}\ }\textbf {\bibinfo {volume} {285}},\
  \bibinfo {pages} {L45} (\bibinfo {year} {1984})}\BibitemShut {NoStop}%
\bibitem [{\citenamefont {Kodama}\ and\ \citenamefont
  {Sasaki}(1986)}]{Kodama:1986fg}%
  \BibitemOpen
  \bibfield  {author} {\bibinfo {author} {\bibfnamefont {H.}~\bibnamefont
  {Kodama}}\ and\ \bibinfo {author} {\bibfnamefont {M.}~\bibnamefont
  {Sasaki}},\ }\href {\doibase 10.1142/S0217751X86000137} {\bibfield  {journal}
  {\bibinfo  {journal} {Int. J. Mod. Phys.}\ }\textbf {\bibinfo {volume}
  {A1}},\ \bibinfo {pages} {265} (\bibinfo {year} {1986})}\BibitemShut
  {NoStop}%
\bibitem [{\citenamefont {Kodama}\ and\ \citenamefont
  {Sasaki}(1987)}]{Kodama:1986ud}%
  \BibitemOpen
  \bibfield  {author} {\bibinfo {author} {\bibfnamefont {H.}~\bibnamefont
  {Kodama}}\ and\ \bibinfo {author} {\bibfnamefont {M.}~\bibnamefont
  {Sasaki}},\ }\href {\doibase 10.1142/S0217751X8700020X} {\bibfield  {journal}
  {\bibinfo  {journal} {Int. J. Mod. Phys.}\ }\textbf {\bibinfo {volume}
  {A2}},\ \bibinfo {pages} {491} (\bibinfo {year} {1987})}\BibitemShut
  {NoStop}%
\bibitem [{\citenamefont {Hu}\ and\ \citenamefont
  {Sugiyama}(1995)}]{Hu:1994jd}%
  \BibitemOpen
  \bibfield  {author} {\bibinfo {author} {\bibfnamefont {W.}~\bibnamefont
  {Hu}}\ and\ \bibinfo {author} {\bibfnamefont {N.}~\bibnamefont {Sugiyama}},\
  }\href {\doibase 10.1103/PhysRevD.51.2599} {\bibfield  {journal} {\bibinfo
  {journal} {Phys. Rev.}\ }\textbf {\bibinfo {volume} {D51}},\ \bibinfo {pages}
  {2599} (\bibinfo {year} {1995})},\ \Eprint
  {http://arxiv.org/abs/astro-ph/9411008} {astro-ph/9411008} \BibitemShut
  {NoStop}%
\bibitem [{\citenamefont {Moodley}\ \emph {et~al.}(2004)\citenamefont
  {Moodley}, \citenamefont {Bucher}, \citenamefont {Dunkley}, \citenamefont
  {Ferreira},\ and\ \citenamefont {Skordis}}]{Moodley:2004nz}%
  \BibitemOpen
  \bibfield  {author} {\bibinfo {author} {\bibfnamefont {K.}~\bibnamefont
  {Moodley}}, \bibinfo {author} {\bibfnamefont {M.}~\bibnamefont {Bucher}},
  \bibinfo {author} {\bibfnamefont {J.}~\bibnamefont {Dunkley}}, \bibinfo
  {author} {\bibfnamefont {P.~G.}\ \bibnamefont {Ferreira}}, \ and\ \bibinfo
  {author} {\bibfnamefont {C.}~\bibnamefont {Skordis}},\ }\href {\doibase
  10.1103/PhysRevD.70.103520} {\bibfield  {journal} {\bibinfo  {journal} {Phys.
  Rev.}\ }\textbf {\bibinfo {volume} {D70}},\ \bibinfo {pages} {103520}
  (\bibinfo {year} {2004})},\ \Eprint {http://arxiv.org/abs/astro-ph/0407304}
  {astro-ph/0407304} \BibitemShut {NoStop}%
\bibitem [{\citenamefont {Bean}\ \emph {et~al.}(2006)\citenamefont {Bean},
  \citenamefont {Dunkley},\ and\ \citenamefont {Pierpaoli}}]{Bean:2006qz}%
  \BibitemOpen
  \bibfield  {author} {\bibinfo {author} {\bibfnamefont {R.}~\bibnamefont
  {Bean}}, \bibinfo {author} {\bibfnamefont {J.}~\bibnamefont {Dunkley}}, \
  and\ \bibinfo {author} {\bibfnamefont {E.}~\bibnamefont {Pierpaoli}},\ }\href
  {\doibase 10.1103/PhysRevD.74.063503} {\bibfield  {journal} {\bibinfo
  {journal} {Phys. Rev.}\ }\textbf {\bibinfo {volume} {D74}},\ \bibinfo {pages}
  {063503} (\bibinfo {year} {2006})},\ \Eprint
  {http://arxiv.org/abs/astro-ph/0606685} {astro-ph/0606685} \BibitemShut
  {NoStop}%
\bibitem [{\citenamefont {Lewis}\ and\ \citenamefont
  {Challinor}(2002)}]{Lewis:2002nc}%
  \BibitemOpen
  \bibfield  {author} {\bibinfo {author} {\bibfnamefont {A.}~\bibnamefont
  {Lewis}}\ and\ \bibinfo {author} {\bibfnamefont {A.}~\bibnamefont
  {Challinor}},\ }\href {\doibase 10.1103/PhysRevD.66.023531} {\bibfield
  {journal} {\bibinfo  {journal} {Phys.Rev.}\ }\textbf {\bibinfo {volume}
  {D66}},\ \bibinfo {pages} {023531} (\bibinfo {year} {2002})},\ \Eprint
  {http://arxiv.org/abs/astro-ph/0203507} {astro-ph/0203507} \BibitemShut
  {NoStop}%
\bibitem [{\citenamefont {Lewis}(2002)}]{cambnotes}%
  \BibitemOpen
  \bibfield  {author} {\bibinfo {author} {\bibfnamefont {A.}~\bibnamefont
  {Lewis}},\ }\href@noop {} {\bibfield  {journal} {\bibinfo  {journal}
  {http://cosmologist.info/notes/CAMB.pdf}\ } (\bibinfo {year}
  {2002})}\BibitemShut {NoStop}%
\bibitem [{\citenamefont {Holder}\ \emph {et~al.}(2010)\citenamefont {Holder},
  \citenamefont {Nollett},\ and\ \citenamefont {van Engelen}}]{Holder:2009gd}%
  \BibitemOpen
  \bibfield  {author} {\bibinfo {author} {\bibfnamefont {G.~P.}\ \bibnamefont
  {Holder}}, \bibinfo {author} {\bibfnamefont {K.~M.}\ \bibnamefont {Nollett}},
  \ and\ \bibinfo {author} {\bibfnamefont {A.}~\bibnamefont {van Engelen}},\
  }\href {\doibase 10.1088/0004-637X/716/2/907} {\bibfield  {journal} {\bibinfo
   {journal} {Astrophys. J.}\ }\textbf {\bibinfo {volume} {716}},\ \bibinfo
  {pages} {907} (\bibinfo {year} {2010})},\ \Eprint
  {http://arxiv.org/abs/0907.3919} {arXiv:0907.3919} \BibitemShut {NoStop}%
\bibitem [{\citenamefont {Gordon}\ and\ \citenamefont
  {Pritchard}(2009)}]{Gordon:2009wx}%
  \BibitemOpen
  \bibfield  {author} {\bibinfo {author} {\bibfnamefont {C.}~\bibnamefont
  {Gordon}}\ and\ \bibinfo {author} {\bibfnamefont {J.~R.}\ \bibnamefont
  {Pritchard}},\ }\href {\doibase 10.1103/PhysRevD.80.063535} {\bibfield
  {journal} {\bibinfo  {journal} {Phys.Rev.}\ }\textbf {\bibinfo {volume}
  {D80}},\ \bibinfo {pages} {063535} (\bibinfo {year} {2009})},\ \Eprint
  {http://arxiv.org/abs/0907.5400} {arXiv:0907.5400} \BibitemShut {NoStop}%
\bibitem [{\citenamefont {De~Simone}\ and\ \citenamefont
  {Kobayashi}(2016)}]{DeSimone:2016ofp}%
  \BibitemOpen
  \bibfield  {author} {\bibinfo {author} {\bibfnamefont {A.}~\bibnamefont
  {De~Simone}}\ and\ \bibinfo {author} {\bibfnamefont {T.}~\bibnamefont
  {Kobayashi}},\ }\href@noop {} {\  (\bibinfo {year} {2016})},\ \Eprint
  {http://arxiv.org/abs/1605.00670} {arXiv:1605.00670} \BibitemShut {NoStop}%
\bibitem [{\citenamefont {Lyth}\ \emph {et~al.}(2003)\citenamefont {Lyth},
  \citenamefont {Ungarelli},\ and\ \citenamefont {Wands}}]{Lyth:2002my}%
  \BibitemOpen
  \bibfield  {author} {\bibinfo {author} {\bibfnamefont {D.~H.}\ \bibnamefont
  {Lyth}}, \bibinfo {author} {\bibfnamefont {C.}~\bibnamefont {Ungarelli}}, \
  and\ \bibinfo {author} {\bibfnamefont {D.}~\bibnamefont {Wands}},\ }\href
  {\doibase 10.1103/PhysRevD.67.023503} {\bibfield  {journal} {\bibinfo
  {journal} {Phys.Rev.}\ }\textbf {\bibinfo {volume} {D67}},\ \bibinfo {pages}
  {023503} (\bibinfo {year} {2003})},\ \Eprint
  {http://arxiv.org/abs/astro-ph/0208055} {astro-ph/0208055} \BibitemShut
  {NoStop}%
\bibitem [{\citenamefont {Gupta}\ \emph {et~al.}(2004)\citenamefont {Gupta},
  \citenamefont {Malik},\ and\ \citenamefont {Wands}}]{Gupta:2003jc}%
  \BibitemOpen
  \bibfield  {author} {\bibinfo {author} {\bibfnamefont {S.}~\bibnamefont
  {Gupta}}, \bibinfo {author} {\bibfnamefont {K.~A.}\ \bibnamefont {Malik}}, \
  and\ \bibinfo {author} {\bibfnamefont {D.}~\bibnamefont {Wands}},\ }\href
  {\doibase 10.1103/PhysRevD.69.063513} {\bibfield  {journal} {\bibinfo
  {journal} {Phys.Rev.}\ }\textbf {\bibinfo {volume} {D69}},\ \bibinfo {pages}
  {063513} (\bibinfo {year} {2004})},\ \Eprint
  {http://arxiv.org/abs/astro-ph/0311562} {astro-ph/0311562} \BibitemShut
  {NoStop}%
\bibitem [{\citenamefont {Gordon}\ and\ \citenamefont
  {Lewis}(2003)}]{Gordon:2002gv}%
  \BibitemOpen
  \bibfield  {author} {\bibinfo {author} {\bibfnamefont {C.}~\bibnamefont
  {Gordon}}\ and\ \bibinfo {author} {\bibfnamefont {A.}~\bibnamefont {Lewis}},\
  }\href {\doibase 10.1103/PhysRevD.67.123513} {\bibfield  {journal} {\bibinfo
  {journal} {Phys.Rev.}\ }\textbf {\bibinfo {volume} {D67}},\ \bibinfo {pages}
  {123513} (\bibinfo {year} {2003})},\ \Eprint
  {http://arxiv.org/abs/astro-ph/0212248} {astro-ph/0212248} \BibitemShut
  {NoStop}%
\bibitem [{\citenamefont {Lyth}\ and\ \citenamefont
  {Wands}(2002)}]{Lyth:2001nq}%
  \BibitemOpen
  \bibfield  {author} {\bibinfo {author} {\bibfnamefont {D.~H.}\ \bibnamefont
  {Lyth}}\ and\ \bibinfo {author} {\bibfnamefont {D.}~\bibnamefont {Wands}},\
  }\href {\doibase 10.1016/S0370-2693(01)01366-1} {\bibfield  {journal}
  {\bibinfo  {journal} {Phys.Lett.}\ }\textbf {\bibinfo {volume} {B524}},\
  \bibinfo {pages} {5} (\bibinfo {year} {2002})},\ \Eprint
  {http://arxiv.org/abs/hep-ph/0110002} {hep-ph/0110002} \BibitemShut {NoStop}%
\bibitem [{\citenamefont {He}\ \emph {et~al.}(2015)\citenamefont {He},
  \citenamefont {Grin},\ and\ \citenamefont {Hu}}]{He:2015msa}%
  \BibitemOpen
  \bibfield  {author} {\bibinfo {author} {\bibfnamefont {C.}~\bibnamefont
  {He}}, \bibinfo {author} {\bibfnamefont {D.}~\bibnamefont {Grin}}, \ and\
  \bibinfo {author} {\bibfnamefont {W.}~\bibnamefont {Hu}},\ }\href {\doibase
  10.1103/PhysRevD.92.063018} {\bibfield  {journal} {\bibinfo  {journal} {Phys.
  Rev.}\ }\textbf {\bibinfo {volume} {D92}},\ \bibinfo {pages} {063018}
  (\bibinfo {year} {2015})},\ \Eprint {http://arxiv.org/abs/1505.00639}
  {arXiv:1505.00639} \BibitemShut {NoStop}%
\bibitem [{\citenamefont {Smith}\ \emph {et~al.}(2017)\citenamefont {Smith},
  \citenamefont {Mu–oz}, \citenamefont {Smith}, \citenamefont {Yee},\ and\
  \citenamefont {Grin}}]{Smith:2017ndr}%
  \BibitemOpen
  \bibfield  {author} {\bibinfo {author} {\bibfnamefont {T.~L.}\ \bibnamefont
  {Smith}}, \bibinfo {author} {\bibfnamefont {J.~B.}\ \bibnamefont {Mu–oz}},
  \bibinfo {author} {\bibfnamefont {R.}~\bibnamefont {Smith}}, \bibinfo
  {author} {\bibfnamefont {K.}~\bibnamefont {Yee}}, \ and\ \bibinfo {author}
  {\bibfnamefont {D.}~\bibnamefont {Grin}},\ }\href@noop {} {\  (\bibinfo
  {year} {2017})},\ \Eprint {http://arxiv.org/abs/1704.03461} {arXiv:1704.03461
  [astro-ph.CO]} \BibitemShut {NoStop}%
\bibitem [{\citenamefont {Grin}\ \emph
  {et~al.}(2011{\natexlab{a}})\citenamefont {Grin}, \citenamefont {Dor\'{e}},\
  and\ \citenamefont {Kamionkowski}}]{Grin:2011nk}%
  \BibitemOpen
  \bibfield  {author} {\bibinfo {author} {\bibfnamefont {D.}~\bibnamefont
  {Grin}}, \bibinfo {author} {\bibfnamefont {O.}~\bibnamefont {Dor\'{e}}}, \
  and\ \bibinfo {author} {\bibfnamefont {M.}~\bibnamefont {Kamionkowski}},\
  }\href {\doibase 10.1103/PhysRevLett.107.261301} {\bibfield  {journal}
  {\bibinfo  {journal} {\prl}\ }\textbf {\bibinfo {volume} {107}},\ \bibinfo
  {pages} {261301} (\bibinfo {year} {2011}{\natexlab{a}})},\ \Eprint
  {http://arxiv.org/abs/1107.1716} {arXiv:1107.1716} \BibitemShut {NoStop}%
\bibitem [{\citenamefont {Grin}\ \emph
  {et~al.}(2011{\natexlab{b}})\citenamefont {Grin}, \citenamefont {Dor\'{e}},\
  and\ \citenamefont {Kamionkowski}}]{Grin:2011tf}%
  \BibitemOpen
  \bibfield  {author} {\bibinfo {author} {\bibfnamefont {D.}~\bibnamefont
  {Grin}}, \bibinfo {author} {\bibfnamefont {O.}~\bibnamefont {Dor\'{e}}}, \
  and\ \bibinfo {author} {\bibfnamefont {M.}~\bibnamefont {Kamionkowski}},\
  }\href {\doibase 10.1103/PhysRevD.84.123003} {\bibfield  {journal} {\bibinfo
  {journal} {Phys.Rev.}\ }\textbf {\bibinfo {volume} {D84}},\ \bibinfo {pages}
  {123003} (\bibinfo {year} {2011}{\natexlab{b}})},\ \Eprint
  {http://arxiv.org/abs/1107.5047} {arXiv:1107.5047} \BibitemShut {NoStop}%
\bibitem [{\citenamefont {Okamoto}\ and\ \citenamefont
  {Hu}(2003)}]{Okamoto:2003zw}%
  \BibitemOpen
  \bibfield  {author} {\bibinfo {author} {\bibfnamefont {T.}~\bibnamefont
  {Okamoto}}\ and\ \bibinfo {author} {\bibfnamefont {W.}~\bibnamefont {Hu}},\
  }\href {\doibase 10.1103/PhysRevD.67.083002} {\bibfield  {journal} {\bibinfo
  {journal} {Phys. Rev.}\ }\textbf {\bibinfo {volume} {D67}},\ \bibinfo {pages}
  {083002} (\bibinfo {year} {2003})},\ \Eprint
  {http://arxiv.org/abs/astro-ph/0301031} {astro-ph/0301031} \BibitemShut
  {NoStop}%
\bibitem [{\citenamefont {Heinrich}\ \emph {et~al.}(2016)\citenamefont
  {Heinrich}, \citenamefont {Grin},\ and\ \citenamefont
  {Hu}}]{Heinrich:2016gqe}%
  \BibitemOpen
  \bibfield  {author} {\bibinfo {author} {\bibfnamefont {C.~H.}\ \bibnamefont
  {Heinrich}}, \bibinfo {author} {\bibfnamefont {D.}~\bibnamefont {Grin}}, \
  and\ \bibinfo {author} {\bibfnamefont {W.}~\bibnamefont {Hu}},\ }\href
  {\doibase 10.1103/PhysRevD.94.043534} {\bibfield  {journal} {\bibinfo
  {journal} {Phys. Rev.}\ }\textbf {\bibinfo {volume} {D94}},\ \bibinfo {pages}
  {043534} (\bibinfo {year} {2016})},\ \Eprint
  {http://arxiv.org/abs/1605.08439} {arXiv:1605.08439 [astro-ph.CO]}
  \BibitemShut {NoStop}%
\bibitem [{\citenamefont {Mollerach}(1990)}]{Mollerach:1989hu}%
  \BibitemOpen
  \bibfield  {author} {\bibinfo {author} {\bibfnamefont {S.}~\bibnamefont
  {Mollerach}},\ }\href {\doibase 10.1103/PhysRevD.42.313} {\bibfield
  {journal} {\bibinfo  {journal} {Phys.Rev.}\ }\textbf {\bibinfo {volume}
  {D42}},\ \bibinfo {pages} {313} (\bibinfo {year} {1990})}\BibitemShut
  {NoStop}%
\bibitem [{\citenamefont {Linde}\ and\ \citenamefont
  {Mukhanov}(1997)}]{Linde:1996gt}%
  \BibitemOpen
  \bibfield  {author} {\bibinfo {author} {\bibfnamefont {A.~D.}\ \bibnamefont
  {Linde}}\ and\ \bibinfo {author} {\bibfnamefont {V.~F.}\ \bibnamefont
  {Mukhanov}},\ }\href {\doibase 10.1103/PhysRevD.56.R535} {\bibfield
  {journal} {\bibinfo  {journal} {Phys.Rev.}\ }\textbf {\bibinfo {volume}
  {D56}},\ \bibinfo {pages} {535} (\bibinfo {year} {1997})},\ \Eprint
  {http://arxiv.org/abs/astro-ph/9610219} {astro-ph/9610219} \BibitemShut
  {NoStop}%
\bibitem [{\citenamefont {Lyth}\ and\ \citenamefont
  {Wands}(2003)}]{Lyth:2003ip}%
  \BibitemOpen
  \bibfield  {author} {\bibinfo {author} {\bibfnamefont {D.~H.}\ \bibnamefont
  {Lyth}}\ and\ \bibinfo {author} {\bibfnamefont {D.}~\bibnamefont {Wands}},\
  }\href {\doibase 10.1103/PhysRevD.68.103516} {\bibfield  {journal} {\bibinfo
  {journal} {Phys.Rev.}\ }\textbf {\bibinfo {volume} {D68}},\ \bibinfo {pages}
  {103516} (\bibinfo {year} {2003})},\ \Eprint
  {http://arxiv.org/abs/astro-ph/0306500} {astro-ph/0306500} \BibitemShut
  {NoStop}%
\bibitem [{\citenamefont {Lemoine}\ and\ \citenamefont
  {Martin}(2007)}]{Lemoine:2006sc}%
  \BibitemOpen
  \bibfield  {author} {\bibinfo {author} {\bibfnamefont {M.}~\bibnamefont
  {Lemoine}}\ and\ \bibinfo {author} {\bibfnamefont {J.}~\bibnamefont
  {Martin}},\ }\href {\doibase 10.1103/PhysRevD.75.063504} {\bibfield
  {journal} {\bibinfo  {journal} {Phys.Rev.}\ }\textbf {\bibinfo {volume}
  {D75}},\ \bibinfo {pages} {063504} (\bibinfo {year} {2007})},\ \Eprint
  {http://arxiv.org/abs/astro-ph/0611948} {astro-ph/0611948} \BibitemShut
  {NoStop}%
\bibitem [{\citenamefont {Ade}\ \emph {et~al.}(2015{\natexlab{c}})\citenamefont
  {Ade} \emph {et~al.}}]{Ade:2015xua}%
  \BibitemOpen
  \bibfield  {author} {\bibinfo {author} {\bibfnamefont {P.~A.~R.}\
  \bibnamefont {Ade}} \emph {et~al.} (\bibinfo {collaboration} {Planck}),\
  }\href@noop {} {\  (\bibinfo {year} {2015}{\natexlab{c}})},\ \Eprint
  {http://arxiv.org/abs/1502.01589} {arXiv:1502.01589} \BibitemShut {NoStop}%
\bibitem [{\citenamefont {Lewis}\ \emph {et~al.}(2000)\citenamefont {Lewis},
  \citenamefont {Challinor},\ and\ \citenamefont {Lasenby}}]{Lewis:1999bs}%
  \BibitemOpen
  \bibfield  {author} {\bibinfo {author} {\bibfnamefont {A.}~\bibnamefont
  {Lewis}}, \bibinfo {author} {\bibfnamefont {A.}~\bibnamefont {Challinor}}, \
  and\ \bibinfo {author} {\bibfnamefont {A.}~\bibnamefont {Lasenby}},\ }\href
  {\doibase 10.1086/309179} {\bibfield  {journal} {\bibinfo  {journal}
  {Astrophys. J.}\ }\textbf {\bibinfo {volume} {538}},\ \bibinfo {pages} {473}
  (\bibinfo {year} {2000})},\ \Eprint {http://arxiv.org/abs/astro-ph/9911177}
  {arXiv:astro-ph/9911177 [astro-ph]} \BibitemShut {NoStop}%
\bibitem [{\citenamefont {Lewis}(2005)}]{Lewis:2005tp}%
  \BibitemOpen
  \bibfield  {author} {\bibinfo {author} {\bibfnamefont {A.}~\bibnamefont
  {Lewis}},\ }\href {\doibase 10.1103/PhysRevD.71.083008} {\bibfield  {journal}
  {\bibinfo  {journal} {Phys. Rev.}\ }\textbf {\bibinfo {volume} {D71}},\
  \bibinfo {pages} {083008} (\bibinfo {year} {2005})},\ \Eprint
  {http://arxiv.org/abs/astro-ph/0502469} {arXiv:astro-ph/0502469 [astro-ph]}
  \BibitemShut {NoStop}%
\bibitem [{\citenamefont {Hamimeche}\ and\ \citenamefont
  {Lewis}(2008)}]{Hamimeche:2008ai}%
  \BibitemOpen
  \bibfield  {author} {\bibinfo {author} {\bibfnamefont {S.}~\bibnamefont
  {Hamimeche}}\ and\ \bibinfo {author} {\bibfnamefont {A.}~\bibnamefont
  {Lewis}},\ }\href {\doibase 10.1103/PhysRevD.77.103013} {\bibfield  {journal}
  {\bibinfo  {journal} {Phys. Rev.}\ }\textbf {\bibinfo {volume} {D77}},\
  \bibinfo {pages} {103013} (\bibinfo {year} {2008})},\ \Eprint
  {http://arxiv.org/abs/0801.0554} {arXiv:0801.0554 [astro-ph]} \BibitemShut
  {NoStop}%
\bibitem [{\citenamefont {Gorski}\ \emph {et~al.}(2005)\citenamefont {Gorski},
  \citenamefont {Hivon}, \citenamefont {Banday}, \citenamefont {Wandelt},
  \citenamefont {Hansen}, \citenamefont {Reinecke},\ and\ \citenamefont
  {Bartelman}}]{Gorski:2004by}%
  \BibitemOpen
  \bibfield  {author} {\bibinfo {author} {\bibfnamefont {K.~M.}\ \bibnamefont
  {Gorski}}, \bibinfo {author} {\bibfnamefont {E.}~\bibnamefont {Hivon}},
  \bibinfo {author} {\bibfnamefont {A.~J.}\ \bibnamefont {Banday}}, \bibinfo
  {author} {\bibfnamefont {B.~D.}\ \bibnamefont {Wandelt}}, \bibinfo {author}
  {\bibfnamefont {F.~K.}\ \bibnamefont {Hansen}}, \bibinfo {author}
  {\bibfnamefont {M.}~\bibnamefont {Reinecke}}, \ and\ \bibinfo {author}
  {\bibfnamefont {M.}~\bibnamefont {Bartelman}},\ }\href {\doibase
  10.1086/427976} {\bibfield  {journal} {\bibinfo  {journal} {Astrophys. J.}\
  }\textbf {\bibinfo {volume} {622}},\ \bibinfo {pages} {759} (\bibinfo {year}
  {2005})},\ \Eprint {http://arxiv.org/abs/astro-ph/0409513}
  {arXiv:astro-ph/0409513 [astro-ph]} \BibitemShut {NoStop}%
\bibitem [{\citenamefont {Lewis}\ \emph {et~al.}(2011)\citenamefont {Lewis},
  \citenamefont {Challinor},\ and\ \citenamefont {Hanson}}]{Lewis:2011fk}%
  \BibitemOpen
  \bibfield  {author} {\bibinfo {author} {\bibfnamefont {A.}~\bibnamefont
  {Lewis}}, \bibinfo {author} {\bibfnamefont {A.}~\bibnamefont {Challinor}}, \
  and\ \bibinfo {author} {\bibfnamefont {D.}~\bibnamefont {Hanson}},\ }\href
  {\doibase 10.1088/1475-7516/2011/03/018} {\bibfield  {journal} {\bibinfo
  {journal} {JCAP}\ }\textbf {\bibinfo {volume} {1103}},\ \bibinfo {pages}
  {018} (\bibinfo {year} {2011})},\ \Eprint {http://arxiv.org/abs/1101.2234}
  {arXiv:1101.2234} \BibitemShut {NoStop}%
\bibitem [{\citenamefont {Zaldarriaga}\ and\ \citenamefont
  {Seljak}(1999)}]{Zaldarriaga:1998te}%
  \BibitemOpen
  \bibfield  {author} {\bibinfo {author} {\bibfnamefont {M.}~\bibnamefont
  {Zaldarriaga}}\ and\ \bibinfo {author} {\bibfnamefont {U.}~\bibnamefont
  {Seljak}},\ }\href {\doibase 10.1103/PhysRevD.59.123507} {\bibfield
  {journal} {\bibinfo  {journal} {Phys. Rev.}\ }\textbf {\bibinfo {volume}
  {D59}},\ \bibinfo {pages} {123507} (\bibinfo {year} {1999})},\ \Eprint
  {http://arxiv.org/abs/astro-ph/9810257} {arXiv:astro-ph/9810257} \BibitemShut
  {NoStop}%
\bibitem [{\citenamefont {Hu}(2001)}]{Hu:2001fa}%
  \BibitemOpen
  \bibfield  {author} {\bibinfo {author} {\bibfnamefont {W.}~\bibnamefont
  {Hu}},\ }\href {\doibase 10.1103/PhysRevD.64.083005} {\bibfield  {journal}
  {\bibinfo  {journal} {Phys. Rev.}\ }\textbf {\bibinfo {volume} {D64}},\
  \bibinfo {pages} {083005} (\bibinfo {year} {2001})},\ \Eprint
  {http://arxiv.org/abs/astro-ph/0105117} {astro-ph/0105117} \BibitemShut
  {NoStop}%
\bibitem [{\citenamefont {Smith}\ and\ \citenamefont
  {Zaldarriaga}(2011)}]{Smith:2006ud}%
  \BibitemOpen
  \bibfield  {author} {\bibinfo {author} {\bibfnamefont {K.~M.}\ \bibnamefont
  {Smith}}\ and\ \bibinfo {author} {\bibfnamefont {M.}~\bibnamefont
  {Zaldarriaga}},\ }\href {\doibase 10.1111/j.1365-2966.2010.18175.x}
  {\bibfield  {journal} {\bibinfo  {journal} {Mon. Not. Roy. Astron. Soc.}\
  }\textbf {\bibinfo {volume} {417}},\ \bibinfo {pages} {2} (\bibinfo {year}
  {2011})},\ \Eprint {http://arxiv.org/abs/astro-ph/0612571} {astro-ph/0612571}
  \BibitemShut {NoStop}%
\bibitem [{\citenamefont {Kim}\ \emph {et~al.}(2013)\citenamefont {Kim},
  \citenamefont {Rotti},\ and\ \citenamefont {Komatsu}}]{Kim:2013nea}%
  \BibitemOpen
  \bibfield  {author} {\bibinfo {author} {\bibfnamefont {J.}~\bibnamefont
  {Kim}}, \bibinfo {author} {\bibfnamefont {A.}~\bibnamefont {Rotti}}, \ and\
  \bibinfo {author} {\bibfnamefont {E.}~\bibnamefont {Komatsu}},\ }\href
  {\doibase 10.1088/1475-7516/2013/04/021} {\bibfield  {journal} {\bibinfo
  {journal} {JCAP}\ }\textbf {\bibinfo {volume} {1304}},\ \bibinfo {pages}
  {021} (\bibinfo {year} {2013})},\ \Eprint {http://arxiv.org/abs/1302.5799}
  {arXiv:1302.5799} \BibitemShut {NoStop}%
\bibitem [{\citenamefont {Ade}\ \emph {et~al.}(2015{\natexlab{d}})\citenamefont
  {Ade} \emph {et~al.}}]{Ade:2015dva}%
  \BibitemOpen
  \bibfield  {author} {\bibinfo {author} {\bibfnamefont {P.~A.~R.}\
  \bibnamefont {Ade}} \emph {et~al.} (\bibinfo {collaboration} {Planck}),\
  }\href@noop {} {\  (\bibinfo {year} {2015}{\natexlab{d}})},\ \Eprint
  {http://arxiv.org/abs/1502.01595} {arXiv:1502.01595} \BibitemShut {NoStop}%
\bibitem [{\citenamefont {Larsen}\ \emph {et~al.}(2016)\citenamefont {Larsen},
  \citenamefont {Challinor}, \citenamefont {Sherwin},\ and\ \citenamefont
  {Mak}}]{Larsen:2016wpa}%
  \BibitemOpen
  \bibfield  {author} {\bibinfo {author} {\bibfnamefont {P.}~\bibnamefont
  {Larsen}}, \bibinfo {author} {\bibfnamefont {A.}~\bibnamefont {Challinor}},
  \bibinfo {author} {\bibfnamefont {B.~D.}\ \bibnamefont {Sherwin}}, \ and\
  \bibinfo {author} {\bibfnamefont {D.}~\bibnamefont {Mak}},\ }\href {\doibase
  10.1103/PhysRevLett.117.151102} {\bibfield  {journal} {\bibinfo  {journal}
  {Phys. Rev. Lett.}\ }\textbf {\bibinfo {volume} {117}},\ \bibinfo {pages}
  {151102} (\bibinfo {year} {2016})},\ \Eprint
  {http://arxiv.org/abs/1607.05733} {arXiv:1607.05733 [astro-ph.CO]}
  \BibitemShut {NoStop}%
\bibitem [{\citenamefont {Carron}\ \emph {et~al.}(2017)\citenamefont {Carron},
  \citenamefont {Lewis},\ and\ \citenamefont {Challinor}}]{Carron:2017vfg}%
  \BibitemOpen
  \bibfield  {author} {\bibinfo {author} {\bibfnamefont {J.}~\bibnamefont
  {Carron}}, \bibinfo {author} {\bibfnamefont {A.}~\bibnamefont {Lewis}}, \
  and\ \bibinfo {author} {\bibfnamefont {A.}~\bibnamefont {Challinor}},\ }\href
  {\doibase 10.1088/1475-7516/2017/05/035} {\bibfield  {journal} {\bibinfo
  {journal} {JCAP}\ }\textbf {\bibinfo {volume} {1705}},\ \bibinfo {pages}
  {035} (\bibinfo {year} {2017})},\ \Eprint {http://arxiv.org/abs/1701.01712}
  {arXiv:1701.01712 [astro-ph.CO]} \BibitemShut {NoStop}%
\bibitem [{\citenamefont {Manzotti}\ \emph {et~al.}(2017)\citenamefont
  {Manzotti} \emph {et~al.}}]{Manzotti:2017net}%
  \BibitemOpen
  \bibfield  {author} {\bibinfo {author} {\bibfnamefont {A.}~\bibnamefont
  {Manzotti}} \emph {et~al.} (\bibinfo {collaboration} {Herschel, SPT}),\
  }\href@noop {} {\  (\bibinfo {year} {2017})},\ \Eprint
  {http://arxiv.org/abs/1701.04396} {arXiv:1701.04396 [astro-ph.CO]}
  \BibitemShut {NoStop}%
\end{thebibliography}%

\end{document}